\documentclass[twocolumn]{aastex6}
\usepackage{graphicx}

\usepackage{placeins}
\usepackage{amsmath}
\newcommand{\degree}{$^{\circ}$}

\newcommand{\kms}{km s$^{-1}$}

\newcommand{\alphafe}{[$\alpha$/Fe]}
\newcommand{\feh}{[Fe/H]}
\newcommand{\mgfe}{[Mg/Fe]}

\begin{document}

\title{Disentangling the Galactic Halo with APOGEE: I. Chemical and Kinematical Investigation of Distinct Metal-Poor Populations}

\author{Christian R. Hayes\altaffilmark{1}, Steven R. Majewski\altaffilmark{1}, Matthew Shetrone\altaffilmark{2}, Emma Fern\'{a}ndez-Alvar\altaffilmark{3}, Carlos Allende Prieto\altaffilmark{4,5}, William J. Schuster\altaffilmark{6}, Leticia Carigi\altaffilmark{3}, Katia Cunha\altaffilmark{7,8}, Verne V. Smith\altaffilmark{9}, Jennifer Sobeck\altaffilmark{10}, Andres Almeida\altaffilmark{11}, Timothy C. Beers\altaffilmark{12}, Ricardo Carrera\altaffilmark{4}, J. G. Fern\'{a}ndez-Trincado\altaffilmark{13,14}, D. A. Garc\'{i}a-Hern\'{a}ndez\altaffilmark{4,5}, Doug Geisler\altaffilmark{13}, Richard R. Lane\altaffilmark{15,16}, Sara Lucatello\altaffilmark{17}, Allison M. Matthews\altaffilmark{1}, Dante Minniti\altaffilmark{18,19,20}, Christian Nitschelm\altaffilmark{21}, Baitian Tang\altaffilmark{13}, Patricia B. Tissera\altaffilmark{18,19}, Olga Zamora\altaffilmark{4,5}}
\altaffiltext{1}{Department of Astronomy, University of Virginia, Charlottesville, VA 22904-4325, USA}
\altaffiltext{2}{University of Texas at Austin, McDonald Observatory, McDonald Observatory, TX 79734, USA}
\altaffiltext{3}{Instituto de Astronom\'{i}a, Universidad Nacional Aut\'{o}noma de M\'{e}xico, Apartado Postal 70-264, M\'{e}xico D.F., 04510 M\'{e}xico}
\altaffiltext{4}{Instituto de Astrof\'{i}sica de Canarias (IAC), V\'{i}a L\'{a}ctea, E-38205 La Laguna, Tenerife, Spain}
\altaffiltext{5}{Departamento de Astrof\'{i}sica, Universidad de La Laguna (ULL), E-38206 La Laguna, Tenerife, Spain}
\altaffiltext{6}{Instituto de Astronom\'{i}a, Universidad Nacional Aut\'{o}noma de M\'{e}xico, AP 106, 22800, Ensenada, B.C., M\'{e}xico}
\altaffiltext{7}{Observat\'{o}rio Nacional, 77 Rua General Jos\'{e} Cristino, Rio de Janeiro, 20921-400, Brazil}
\altaffiltext{8}{Steward Observatory, University of Arizona, 933 North Cherry Avenue, Tucson, AZ 85721, USA}
\altaffiltext{9}{National Optical Astronomy Observatories, Tucson, AZ, 85719, USA}
\altaffiltext{10}{Department of Astronomy, Box 351580, University of Washington, Seattle, WA 98195, USA}
\altaffiltext{11}{Instituto de Investigaci\'{o}n Multidisciplinario en Ciencia y Tecnolog\'{i}a, Universidad de La Serena, Benavente 980, La Serena, Chile}
\altaffiltext{12}{Department of Physics and JINA Center for the Evolution of the Elements, University of Notre Dame, Notre Dame, IN 46556, USA}
\altaffiltext{13}{Departamento de Astronom\'{i}a, Universidad de Concepci\'{o}n, Casilla 160-C, Concepci\'{o}n, Chile}
\altaffiltext{14}{Institut Utinam, CNRS UMR6213, Univ. Bourgogne Franche-Comt\'e, OSU THETA , Observatoire de Besan\c{c}on, BP 1615, 25010 Besan\c{c}on Cedex, France}
\altaffiltext{15}{Millennium Institute of Astrophysics, Av. Vicu\~{n}a Mackenna 4860, 782-0436 Macul, Santiago, Chile}
\altaffiltext{16}{Instituto de Astrof\'{i}sica, Pontificia Universidad Cat\'{o}lica de Chile, Av. Vicu\~{n}a Mackenna 4860, 782-0436 Macul, Santiago, Chile}
\altaffiltext{17}{Osservatorio Astronomico di Padova $-$ INAF, Vicolo dell'Osservatorio 5, I-35122, Padova, Italy}
\altaffiltext{18}{Departamento de Ciencias F\'{i}sicas, Facultad de Ciencias Exactas, Universidad  Andr\'{e}s Bello,  Av.  Fern\'{a}ndez Concha  700, Las  Condes,  Santiago, Chile}
\altaffiltext{19}{Instituto Milenio de Astrof\'{i}sica, Av. Vicuna Mackenna 4860, Macul, Santiago, Chile}
\altaffiltext{20}{Vatican  Observatory,   V00120  Vatican  City  State, Italy}
\altaffiltext{21}{Unidad de Astronom\'{i}a, Universidad de Antofagasta, Avenida Angamos 601, Antofagasta 1270300, Chile}
\email{crh7gs@virginia.edu}

\begin{abstract}

We find two chemically distinct populations separated relatively cleanly in the [Fe/H] - [Mg/Fe] plane, but also distinguished in other chemical planes, among metal-poor stars (primarily with metallicities [Fe/H] $< -0.9$) observed by the Apache Point Observatory Galactic Evolution Experiment (APOGEE) and analyzed for Data Release 13 (DR13) of the Sloan Digital Sky Survey.  These two stellar populations show the most significant differences in their [X/Fe] ratios for the $\alpha$-elements, C+N, Al, and Ni.  In addition to these populations having differing chemistry, the low metallicity high-Mg population (which we denote the HMg population) exhibits a significant net Galactic rotation, whereas the low-Mg population (or LMg population) has halo-like kinematics with little to no net rotation.  Based on its properties, the origin of the LMg population is likely as an accreted population of stars.  The HMg population shows chemistry (and to an extent kinematics) similar to the thick disk, and is likely associated with {\it in situ} formation.  The distinction between the LMg and HMg populations mimics the differences between the populations of low- and high-$\alpha$ halo stars found in previous studies, suggesting that these are samples of the same two populations.


\end{abstract}


\section{Introduction}
	
A key step to reconstructing the history of the Milky Way's formation and evolution is to identify and characterize its constituent stellar populations. Metal-poor stars probe the early evolution of the Galaxy and give clues to the origin of its first components.  Among the Milky Way (MW) components containing a large fraction of metal-poor stars are the thick disk \citep[originally known as Intermediate Pop II stars and later reidentified by][]{yos82,gr83} via its metal-poor extension \citep[the metal-weak thick disk, MWTD,][]{mor93,cb00,bee02}, globular clusters and dwarf MW satellite galaxies, and the halo, possibly separating into an inner- and outer-halo components \citep{hart87,slz90,allen91,kin94,nor94,car07,car10,bee12}, but containing sub-populations of globular clusters \citep{zinn93} and fields stars accreted from  hierarchical formation, which undoubtedly played a key role in  the formation of the halo.  A long standing problem is whether and how these different populations may be discriminated from one another by their spatial, kinematical and chemical distributions.

A commonly-used strategy is to rely on kinematical definitions to separate stars into populations \citep{venn04,red06,ruc11,ish12,ish13}.  Unfortunately, this is fraught with several difficulties, not least that it requires that the necessary kinematical data are in hand and of sufficient quality to provide meaningful discrimination.  More problematical than these practical requirements is that the kinematical distributions of these various Galactic components typically overlap, so that it is generally not possible to undertake definitive separations of stars into their respective populations with kinematical information alone. Even resorting to simple statistical prescriptions can be perilous given uncertainties in critical priors used to define and fit distribution functions, such as the number of components to fit \citep[see the discussion in][]{car10} and their intrinsic shapes (not necessarily Gaussian) and therefore free parameters.

Nonetheless, studies of the detailed chemistry of kinematically-defined populations have successfully revealed some of the primary chemical characteristics of these metal-poor populations.  The chemical properties of the thick disk and at least some subset of the halo, although not always cleanly distinct but showing overlaps, have been shown to exhibit demonstrably different mean chemistry for numerous chemical elements \citep{ns97,ns10,ns11,ish12,ram12,ish13}.  For example, these studies have shown that at least some fraction of halo stars have lower abundances of $\alpha$-elements (particularly O, Mg, and Si), Na, Ni, Cu, and Zn and higher Eu enrichement than those of the thick disk at metallicities \feh $\ \gtrsim -1.5$.

One early study of the detailed chemical abundances of 29 metal-poor stars suggested that there may be two chemical abundance patterns amongst halo stars, one of which differed from the thick disk abundance pattern \citep{ns97}.  In a subsequent study of an enlarged sample of 94 stars with metallicities $-1.6 < $ [Fe/H] $ -0.4$, \citet{ns10} used chemical abundances to resolve two rather distinct and mostly non-overlapping populations of stars with halo-like kinematics in the \mgfe-\feh \ chemical plane, with one population having chemistry consistent with the thick disk and the other distinctly less Mg-enriched.  Among differences in Mg and other $\alpha$-elements (distinguishable as ``high and low-$\alpha$'' halo star groupings), these two metal-poor populations were shown to have different abundances in many of the odd-$Z$ and heavier elements listed above \citep{ns10,ns11,nav11,ram12,she12,jj14,haw15}.

Furthermore, using isochrone fits to stellar surface temperatures and gravities, these two [$\alpha$/Fe] groupings were shown to exhibit different mean ages, with the low-$\alpha$ population being younger \citep{sch12}.  From the $\alpha$-element abundances and kinematics of the two populations, these past studies have suggested that the low-$\alpha$ population has been accreted through the mergers of dwarf spheroidal-like galaxies \citep[an origin also suggested for ``young halo'' globular clusters; see][]{zinn93}, whereas the high-$\alpha$ stars were likely formed {\it in situ} or have been kicked out from the disk \citep{she12,jon16}.  Recent studies have also revealed low-$\alpha$ bulge stars, most of which are thought to be chemically associated with the thin disk \citep{rec17}.  While most of these low-$\alpha$ bulge stars have $\alpha$-element abundances that seem too high to be associated with the low-$\alpha$ halo population, a few of these ``bulge stars'' may have chemical abundances more similar to the low-$\alpha$ halo population.  It would not be surprising if low-$\alpha$ halo stars were found in the bulge, since the densities of other stellar populations increase toward the center of the Galaxy.


Despite the proven utility of high precision, high resolution spectroscopic measurements of chemical abundances to distinguish chemically distinct populations of metal-poor stars, such work is observationally expensive.  Consequently, previous sample sizes have generally been limited to a few hundred metal-poor stars (as in the references above).  However, the advent of systematic high resolution surveys, such as the APOGEE survey \citep[Apache Point Observatory Galactic Evolution Experiment;][]{maj15}, the Gaia-ESO Survey \citep{gaiaeso}, and the GALAH survey \citep[Galactic Archaeology with HERMES;][]{galah}, brings the opportunity to put these types of studies on a much firmer statistical footing.  In this work, we use data from the APOGEE survey to gain a more comprehensive view of the chemical differences between populations of metal-poor stars.  

The APOGEE-1 survey \citep{maj15} observed $\sim 146,000$ stars with good quality ($S/N \ge 100$), high resolution ($R \sim 22,500$), infrared (1.5-1.7$\mu$m) spectra from which abundances have been derived for up to 23 elemental species in Data Release 13 \citep[DR13;][]{dr13}, at least for more metal-rich and cool stars \citep{hol15}.  Because metal-poor stars are relatively rare and APOGEE, for the most part, uses no special pre-selection for them, they comprise a relatively small fraction of the APOGEE sample.  Nevertheless, the APOGEE-1 sample (according to abundances derived for DR13) includes over 1,000 metal-poor stars having \feh $<-1.0$ extending down to \feh \ $\sim -2.0$ (i.e., a sample several times larger than previous studies) and with reliable chemical abundances for as many as 12 elements.  Such a large sample of metal-poor stars and a highly dimensional chemical space enables robust searches for chemically distinct metal-poor populations, and  allows testing of previous claims with a larger statistical footing.  Moreover, because the main APOGEE survey targets are only selected photometrically, APOGEE-based studies are free of kinematical biases and include stars from a much larger volume of the Galaxy than previous studies, especially those restricted to observing nearby stars with measured proper motions \citep[e.g.,][]{red06,ns10,ns11,ish12,ish13,ben14}.

This work differs from the past study of metal-poor field stars with APOGEE by \citep{haw15} in the lack of any kinematical selection and use of data driven chemical identification of distinct chemical populations that is supported by independent statistical clustering analyses.  In addition, we use APOGEE data from DR13, which, through improvements to the data reduction, stellar parameter, and chemical abundance pipelines has improved APOGEE's chemical abundances and provided a much larger sample of metal-poor stars with accurate chemical abundances compared to that provided by DR12, used by \citep{haw15}.  The DR13 improvements to chemical abundance measurements in particular allow us greater power to statistically discriminate and characterize the  population of proposed low-$\alpha$ accreted halo stars noticed in previous studies from the population of metal-poor stars having higher $\alpha$-element abundances.  

We show that the [Mg/Fe]-[Fe/H] chemical plane is an especially powerful and reliable diagnostic for this population analysis, and one readily provided by APOGEE for the majority of stars, while other element ratios, like [Al/Fe] and [(C+N)/Fe] are equally discriminating, if less available for all stars ([(C+N)/Fe] becomes uncertain at the lowest metallicities in our study; see Section 3.2 for a discussion of the limitations of the C and N abundances).  We provide evidence supporting our new selection criteria in these and other chemical dimensions by presenting the results of multi-dimensional clustering algorithms on the APOGEE-observed metal-poor stars.  Moreover, because our sample is kinematically unbiased, we can more reliably explore and characterize the kinematical properties of these chemical groupings; we show that the two primary [Mg/Fe]-based metal-poor groupings have decidedly different kinematical properties that give clues to their origin and relation to the main spatio-kinematical populations of the Galaxy.

In particular, as suggested by previous work, the high-Mg  population is relatively kinematically cold and rotating while the low-Mg population has hot kinematics consistent with expectations for an accreted population.  Finally, because metal-poor stars characterized by low-$\alpha$ abundance patterns are traditionally attributed to satellite accretion, we compare the detailed chemical properties of our Mg-populations to those in MW satellites.

This paper is organized as follows.  In Section 2 we discuss the data and selection criteria employed to create the parent stellar sample used throughout the paper.  In Section 3 we first discuss our identification of two populations of metal-poor stars based on their \mgfe.  We then examine the chemical signatures of these populations in other dimensions of APOGEE-observed chemical space, and apply multidimensional clustering algorithms to justify further our characterization of these metal-poor populations.  Section 3 also presents the kinematical properties of these populations derived from APOGEE radial velocity data.  In Section 4, we compare our sample of stars and the populations we have defined to those suggested and discussed in previous studies.  We also comment on the possible origins of these populations, aided by a comparison of our data to the abundance patterns of MW satellites.  We present our conclusions in Section 5.  A companion paper, \citet{paper2}, further explores the chemical evolution and star formation histories of the two populations discriminated in this work.

\section{Data}

Using the Sloan 2.5-m telescope at Apache Point Observatory \citep{gun06} as a part of the third installment of the Sloan Digital Sky Survey \citep[SDSS-III,][]{sdss3}, APOGEE spectroscopically observed a relatively homogenous sample of $\sim$ 146,000 MW stars to survey its multiple structural components.  The details of the APOGEE instrument, survey, data, and calibration are outlined in \citet{maj15} and references therein.  Here we present an analysis of the APOGEE data presented in the SDSS Data Release 13 \citep[DR13;][]{dr13}, the first data release of SDSS-IV \citep{sdss4}.  In this data release, a re-analysis of the spectra presented in SDSS DR12 was performed to improve the quality of derived parameters.  Target selection and data reduction  for APOGEE are described in detail by \citet{zas13} and \citet{dln15}, respectively, and the APOGEE Stellar Parameter and Chemical Abundances Pipeline \citep[ASPCAP; for a detailed description see][]{gp15} was used to determine the stellar parameters and chemical abundances from the best fits to pre-computed libraries of synthetic stellar spectra \citep[e.g.,][]{zam15}.

We restrict our analysis to a subsample of stars selected on the basis of a series of APOGEE flags and other constraints.  We first removed any stars with the \textsc{starflags} \textsc{bad\_pixels}, \textsc{very\_bright\_neighbor}, or \textsc{low\_snr} flags set.  We also cut any stars with the following \textsc{aspcapflags}:  \textsc{metals\_warn}, \textsc{rotation\_warn}, \textsc{metals\_bad}, \textsc{star\_bad}, \textsc{rotation\_bad}, or \textsc{sn\_bad}.  Descriptions of these flags can be found online in the SDSS DR13 bitmask documentation\footnote{\url{http://www.sdss.org/dr13/algorithms/bitmasks/}}.

In addition to trimming the APOGEE data set using quality flags, we also require that the visit-to-visit velocity scatter be small, i.e. $V_{scatter} \leq 1$ \kms, because a larger velocity scatter may indicate surface activity, the presence of a companion, or other astrophysical complications that may make determined parameters and abundances less reliable.  Similarly, we only use stars with velocity uncertainties $V_{err} \leq 0.2$ \kms, to exclude stars with large velocity uncertainties that may have less reliably derived parameters.  We have also restricted our analysis to stars with surface temperatures $4000$ K $< T_{\rm eff} < 5500$ K (selected before applying the post-calibration corrections to produce the surface temperatures and gravities listed in Table \ref{table1}), given that, as of DR13, ASPCAP is not yet tuned for reliable parameter estimation for cooler stars or those that are warmer and have weaker lines.

Finally, because we will be primarily concerned with magnesium abundances, we only select stars with $\sigma$[Mg/Fe] $< 0.1$ and S/N $> 100$.  However, in considering other elemental abundances throughout the remainder of the paper, we only examine (but do not remove from the sample) stars with uncertainties on those abundances below 0.1 dex as well.  Because globular clusters are known to exhibit high levels of self-enrichment \citep{gra04}, we have excluded those cluster stars that are easily distinguished from field stars and that can be associated with specific globular clusters (based on proximity, radial velocities, and metallicities).  The Sagittarius dwarf spheroidal galaxy (Sgr dSph) is the only dSph present in DR13, so in addition to removing globular cluster members, we also removed known Sgr dSph members.  While the coolest, most luminous Sgr dSph giants are removed by our $T_{\rm eff} > 4000$ K requirement, we also removed any stars with the \textsc{targflag} \textsc{apogee\_sgr\_dsph}, which was assigned to known Sgr dSph members.  After all of these quality cuts and the exclusion of globular cluster and Sgr dSph stars, we are left with 61,742 stars for study using the calibrated ASPCAP stellar parameters and chemical abundances.

\section{Results and Analysis}

\subsection{Two Populations Seen in [Mg/Fe] Ratios of Metal-Poor Stars}

Visual inspection of the elemental abundances observed by APOGEE for metal-poor stars revealed the most apparent and distinct bimodality in the distribution of \mgfe.  We therefore first present and examine the distribution of Mg abundances.  This strategy is also motivated by previous studies of metal-poor stars that also found two distinct metal-poor groups based on their \alphafe \ ratios, or specifically their \mgfe \ ratios (see Section 1).

\begin{figure}
  \centering
  \includegraphics[scale=0.4]{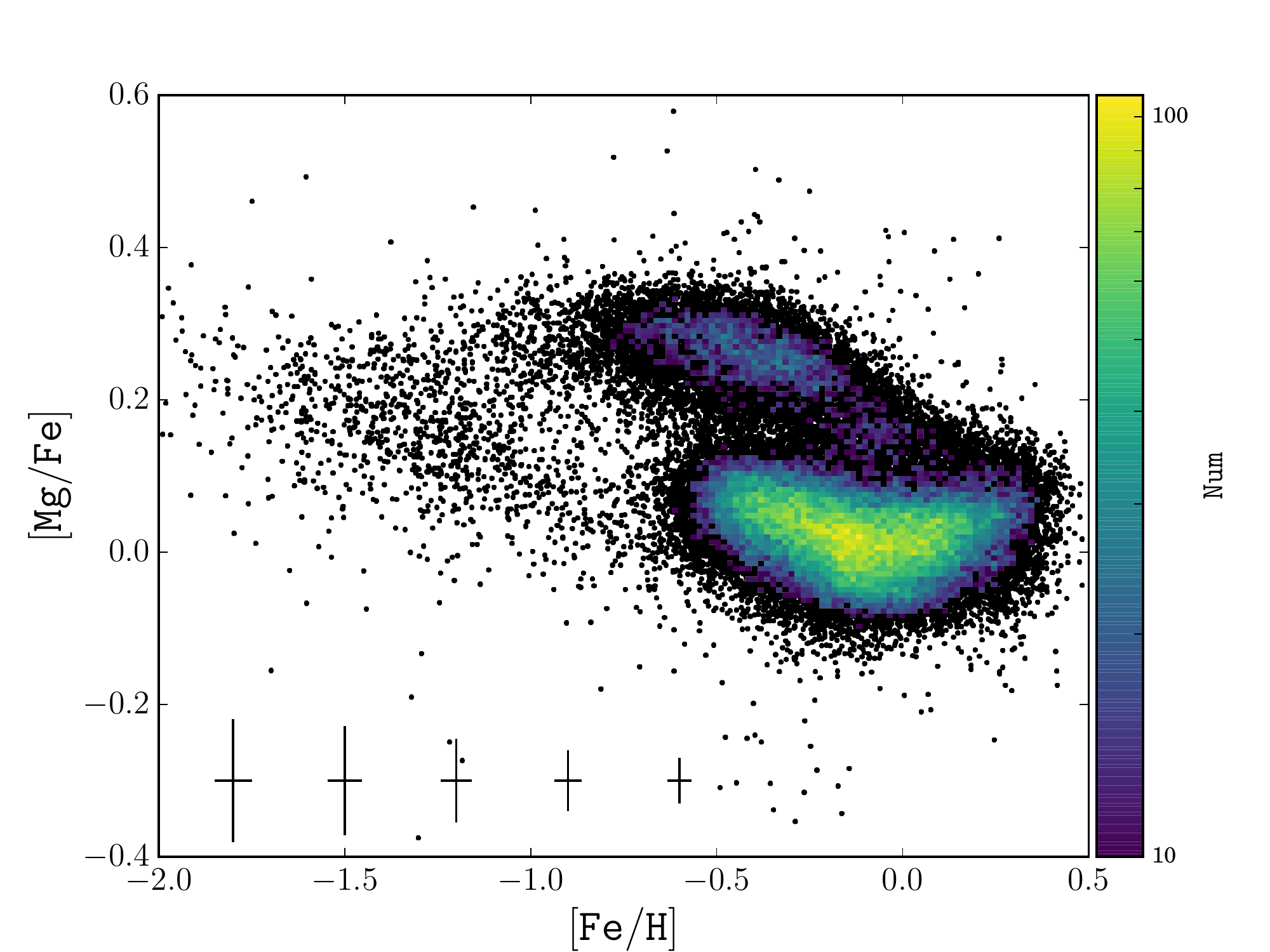}
  \caption{Distribution of [Mg/Fe] with metallicity for the APOGEE DR 13 stars surviving the quality cuts discussed in Section 2.  A 2D histogram is plotted for the highly populated portions of this chemical space (corresponding to the chemical domain of the relatively more metal-rich thin and thick disk populations), while individual stars are plotted where APOGEE observed stars are less populous in this plane.  The plotted error bars show the median abundance uncertainties in 0.3 dex wide metallicity bins.  In addition to the traditional thick disk sequence seen at [Fe/H] $> -1.0$, a noticeable third sequence of stars appears between the low metallicity end of our sample and [Fe/H] $\sim -1.0$, with decreasing [Mg/Fe] with increasing metallicity. }
  \label{mgfe}
\end{figure}

\begin{figure}
  \centering
  \includegraphics[scale=0.4]{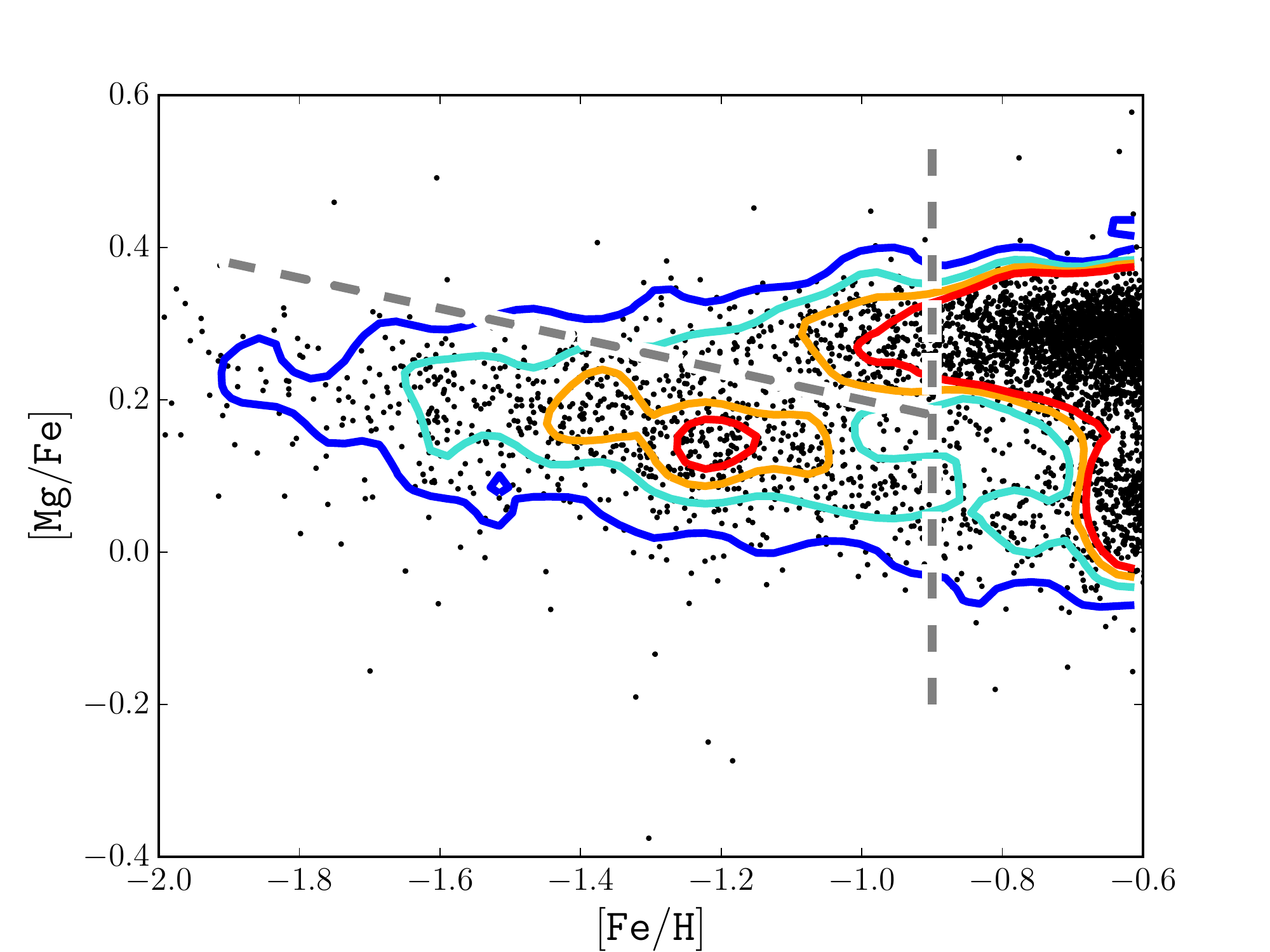}
  \caption{A magnified portion of the metal-poor region of Figure \ref{mgfe}, with contours showing the density of stars in the metal-poor regions of Figure \ref{mgfe}.  The contours are at 5, 15, 25 and 35 stars per 0.0039 dex$^2$.  These contours demonstrate that there is a low density valley separating two higher density regions, one with lower [Mg/Fe], and one with higher [Mg/Fe] that appears to be a metal-poor extension of the thick disk locus.  The sloping dashed line is adopted to separate the two populations based on their [Mg/Fe] and metallicity in Section 3.1. }
  \label{contour}
\end{figure}

Figure \ref{mgfe} shows the distribution of [Mg/Fe] with [Fe/H] for all stars that made it through the quality criteria discussed in Section 2.  Most obvious in this plot are the high- and low-[Mg/Fe] tracks between [Fe/H] $\sim -0.9$ and $+0.4$ nominally corresponding to the thick and thin disks respectively.    In the APOGEE DR 13 data, the high-$\alpha$ sequence commonly associated with the thick disk \citep[e.g.,][and references therein]{bovy16} seems to taper off, and there appears to be a gap between the thick disk and another set of stars with not only a lower level of [Mg/Fe], but decreasing [Mg/Fe] with increasing metallicity.  

This gap is made more apparent in Figure \ref{contour}, where we plot density contours over our data to demonstrate that there is a true low density valley separating the two sequences of metal-poor stars.   The peak-to-valley ratio between the density along the Mg-poor sequence and the density in the valley, tracked by the sloped dashed line in Figure \ref{contour}, gives us an idea of the significance of this second, low-Mg abundance sequence.  At a metallicity of \feh \ $\sim -1.5$ the peak-to-valley ratio is about 1.5, increasing to 2.5 at a metallicity of \feh \ $\sim -1.2$, and it is highest at about 3.0 around a metallicity of $-1.0$.  This indicates that the valley separating these two sequences is more significant at higher metallicities, where the chemical separation between this sequence and the nominal thick disk is larger.  It also highlights that at lower metallicities the two low metallicity sequences overlap more, so that it is more difficult to separate them. 

\begin{figure}
  \centering
  \includegraphics[scale=0.4]{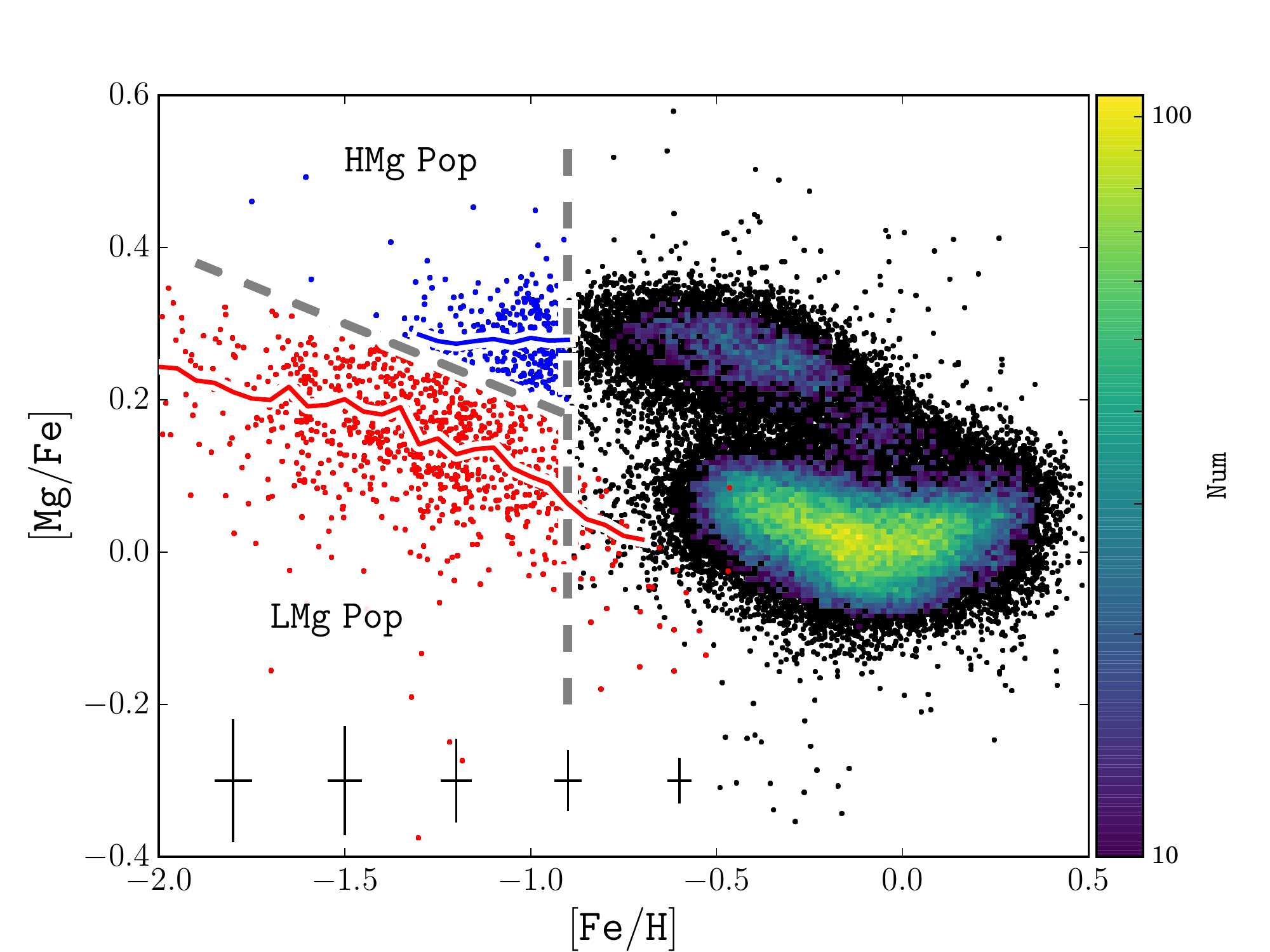}
  \caption{Same as in Figure \ref{mgfe}, but with the initial division to separate the relatively Mg-poor population LMg (red) and the Mg-rich population HMg (blue).  Stars in the LMg population with metallicities [Fe/H] $ > -0.9$ have been selected based on their [(C+N)/Fe] abundance ratios, since they appear to follow the abundance pattern of the LMg population as discussed in Section 3.2.  Over-plotted are the roving boxcar medians of 50 nearest neighbors, again color-coded by population}
  \label{mgfegold}
\end{figure}

Thus, at lower metallicities ([Fe/H] $\lesssim -0.9$) the APOGEE data strongly suggest the existence of two populations of stars chemically differentiated by their [Mg/Fe] patterns.  For simplicity we initially separate the two populations along the gap or valley by [Mg/Fe] $= -0.2 \times $[Fe/H], as shown by the sloped dashed line in Figure \ref{contour}.  We designate the low-[Mg/Fe] population as the Low-Mg (LMg) population and the high-[Mg/Fe] magnesium population as the High-Mg (HMg) population.  For this analysis, we initially restrict our analysis to metallicities of [Fe/H] $\leq -0.9$, to avoid contaminating the LMg population with stars from the thin disk locus.  Note, however, that we show below in Section 3.2 that the LMg population extends to slightly higher metallicities as seen by the consideration of [(C+N)/Fe] ratios.  Our initial division of the LMg and HMg for [Fe/H] $\leq -0.9$ is shown in Figure \ref{mgfegold}.  In Table \ref{table1} we report the relevant properties, stellar parameters, and chemical abundances of the stars categorized into these two populations.

Given that the gap between these populations is not completely devoid of stars, there is some uncertainty in separating them, and there is likely to be some spread of each population across the adopted division, whether due to intrinsic scatter of the true underlying populations or to measurement uncertainties, leading to some low level of cross-contamination that appears to become more significant at lower metallicities ([Fe/H] $\sim -1.3$) where the sequences begin to merge. This is examined in more detail using the full chemical profiles of these populations and clustering algorithms in Section 3.3.  Because the HMg population smoothly connects with the thick disk locus, this population is likely the metal-weak extension of the thick disk.  The origin of the LMg population is not immediately clear and is examined in more detail in later sections.

\begin{deluxetable*}{c l p{4.5cm} c l p{4.5cm}}
\tabletypesize{\scriptsize}
\tablewidth{0pt}
\tablecolumns{3}
\tablecaption{Properties, Parameters, and Population Identification of APOGEE DR13 Metal-Poor Stars \label{table1}}
\tablehead{\colhead{Column} & \colhead{Column Label} & \colhead{Column Description} & \colhead{Column} & \colhead{Column Label} & \colhead{Column Description}}
\startdata
1 & APOGEE\_ID & APOGEE Star ID & 24 & O\_FE\_ERR & Uncertainty on [O/Fe] \\
2 & RA & Right Ascension (decimal degrees) & 25 & MG\_FE & [Mg/Fe] \\
3 & DEC & Declination (decimal degrees) & 26 & MG\_FE\_ERR & Uncertainty on [Mg/Fe] \\
4 & GLON & Galactic Longitude (decimal degrees) & 27 & AL\_FE & [Al/Fe] \\
5 & GLAT & Galactic Latitude (decimal degrees) & 28 & AL\_FE\_ERR & Uncertainty on [Al/Fe] \\
6 & J & 2MASS J magnitude & 29 & SI\_FE & [Si/Fe] \\
7 & H & 2MASS H magnitude & 30 & SI\_FE\_ERR & Uncertainty on [Si/Fe] \\
8 & K & 2MASS Ks magnitude & 31 & K\_FE & [K/Fe] \\
9 & V\_HELIO & Heliocentric radial velocity (km/s) & 32 & K\_FE\_ERR & Uncertainty on [K/Fe] \\
10 & V\_ERR & Radial velocity uncertainty (km/s) & 33 & CA\_FE & [Ca/Fe] \\
11 & TEFF\tablenotemark{a} & Effective surface temperature & 34 & CA\_FE\_ERR & Uncertainty on [Ca/Fe] \\
12 & LOGG\tablenotemark{a} & Surface gravity & 35 & CR\_FE & [Cr/Fe] \\
13 & VMICRO & Microturbulent velocity (km/s) & 36 & CR\_FE\_ERR & Uncertainty on [Cr/Fe] \\
14 & VMACRO & Macroturbulent velocity (km/s) & 37 & MN\_FE & [Mn/Fe]\\
15 & FE\_H & [Fe/H] & 38 & MN\_FE\_ERR & Uncertainty on [Mn/Fe] \\
16 & FE\_H\_ERR & Uncertainty on [Fe/H] & 39 & NI\_FE & [Ni/Fe] \\
17 & ALPHA\_FE & [$\alpha$/Fe] (see text for details) & 40 & NI\_FE\_ERR & Uncertainty on [Ni/Fe] \\
18 & ALPHA\_FE\_ERR & Uncertainty on [$\alpha$/Fe] & 41 & CN\_FE & [(C$+$N)/Fe] \\
19 & C\_FE & [C/Fe] & 42 & CN\_FE\_ERR & Uncertainty on [(C$+$N)/Fe]\\
20 & C\_FE\_ERR & Uncertainty on [C/Fe] & 43 & CNO\_FE & [(C$+$N$+$O)/Fe] \\
21 & N\_FE & [N/Fe] & 44 & CNO\_FE\_ERR & Uncertainty on [(C$+$N$+$O)/Fe] \\
22 & N\_FE\_ERR & Uncertainty on [N/Fe] & 45 & POP\_ID & Assigned Population (LMg or HMg) \\
23 & O\_FE & [O/Fe] & & & \\
\enddata
\tablenotetext{a}{Corrected according to the recommendations in the APOGEE DR13 documentation (\url{http://www.sdss.org/dr13/irspec/parameters/}) to remove surface temperature and gravity trends found post-calibration}
\tablecomments{Table 1 is published in its entirety in the machine-readable format.
      A portion is shown here for guidance regarding its form and content.}
\tablecomments{Null table entries are given values of -9999.}
\end{deluxetable*}

\subsection{Chemical Signatures}

While we have identified a potential division in populations with low metallicities in \mgfe-\feh \ space, for it to have astrophysical significance, we expect it should be revealed in additional dimensions, which we now examine.  In the following subsections we present and analyze the chemical distributions of our sample in other element planes, restricting our analysis of each element to the stars with uncertainties $\sigma$[X/Fe] $< 0.1$ dex in that element alone.

\begin{figure*}
  \centering
  \includegraphics[scale=0.4,trim = 0mm 10mm 0mm 5mm, clip]{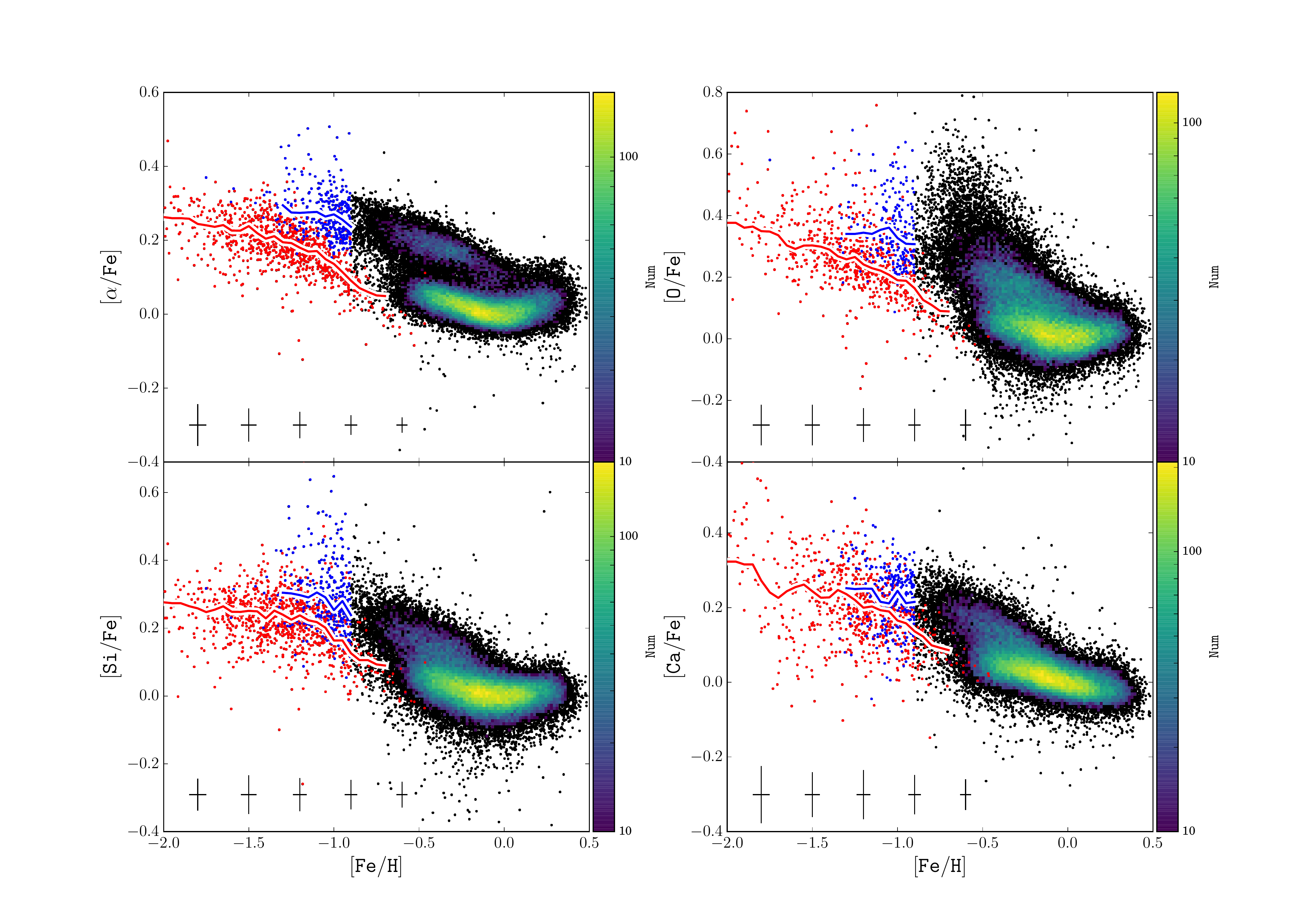}
  \caption{Distribution of different $\alpha$-elements with [Fe/H], with a 2D histogram used for the densely populated regions.  Stars of the LMg and HMg populations are color-coded the same as in Figure \ref{mgfegold}.  Over-plotted are lines of moving medians (using the 50 nearest neighbors), color-coded by population.  The separation between the LMg and HMg populations is smaller in these other $\alpha$-elements than for Mg, but still appears to exist for most of these chemical planes, except potentially that for Ca, where the metal-poor population overlap is greatest. }
  \label{alphagold}
\end{figure*}

\subsubsection{$\alpha$-Elements:  O, Si, and Ca}

We first examine other $\alpha$-elements measured by APOGEE with high precision:  O, Si, and Ca, as well as the ASPCAP global $\alpha$-element parameter \citep[derived from the initial ASPCAP fit to all $\alpha$-elements, O, Mg, Si, Ca, S, Ti; see ][]{hol15}, whose abundances relative to Fe are shown in Figure \ref{alphagold}.  Ti is a commonly studied $\alpha$-element and is measured by APOGEE, however, it is considered unreliable because it is not able to reproduce the [Ti/Fe] bimodality seen in solar neighborhood studies or in other $\alpha$-element abundances measured by APOGEE \citep{hol15}.  The inconsistency with optically derived Ti abundances may be due to the ASPCAP inclusion of lines affected by NLTE or saturation \citep{haw16} in the measurement of Ti, or a high $T_{\rm eff}$ sensitivity of $H$-band TiI lines \citep{sou16}, regardless of the cause, because of this unreliability, we do not analyze Ti here.

Reassuringly, we find that the LMg population is consistently lower in O, Si, and Ca abundances and the HMg population is higher, as seen in magnesium, however, the separation between the LMg and HMg populations is not as clean in these other $\alpha$-elements as it is in magnesium.  The potential exception to this is perhaps in the total $\alpha$, which could be a result of the Mg influence in determining the global $\alpha$-element abundance by ASPCAP.  For the other $\alpha$-element chemical planes, the separation appears to be largest in O, weaker in Si, and weakest in Ca.  

While abundance uncertainties may help obscure differences in $\alpha$-element abundance trends between the LMg and HMg populations, the typical measurement uncertainties for each of the $\alpha$-elements are quite similar (at a given metallicity).  Thus the larger separation in lighter $\alpha$-elements than heavier ones seems to be from an astrophysical source rather than due to differences in random uncertainties (although systematic errors could still obscure differences).  The size of the separation of the LMg and HMg populations in different $\alpha$-elements likely arises from the influence of different types of supernovae.  For example, the LMg population may have experienced enrichment from a higher fraction Type Ia supernovae.  \citet{tsu95} have shown that, while Type Ia supernovae have contributed a negligible amount of O and Mg (about 1\% each) in the solar neighborhood, they have contributed a larger fraction of Si (17\%), Ca (25\%), and Fe (57\%).  They show that these fractions increase in lower mass/metallicity system, such as the LMC, in which Type Ia supernovae still contribute a small fraction of O and Mg (3\%), but make up an even larger fraction of the Ca (44\%) and Fe (76\%).  Another possible explanation for the different separations in $\alpha$-elements is that the two metal-poor populations could have been chemically enriched by supernovae of considerably different progenitor masses or metallicities, which may effect their chemical abundance patterns \citep[see][]{nom13}. 

\begin{figure}
  \centering
  \includegraphics[scale=0.4,trim = 0mm 30mm 0mm 22mm, clip]{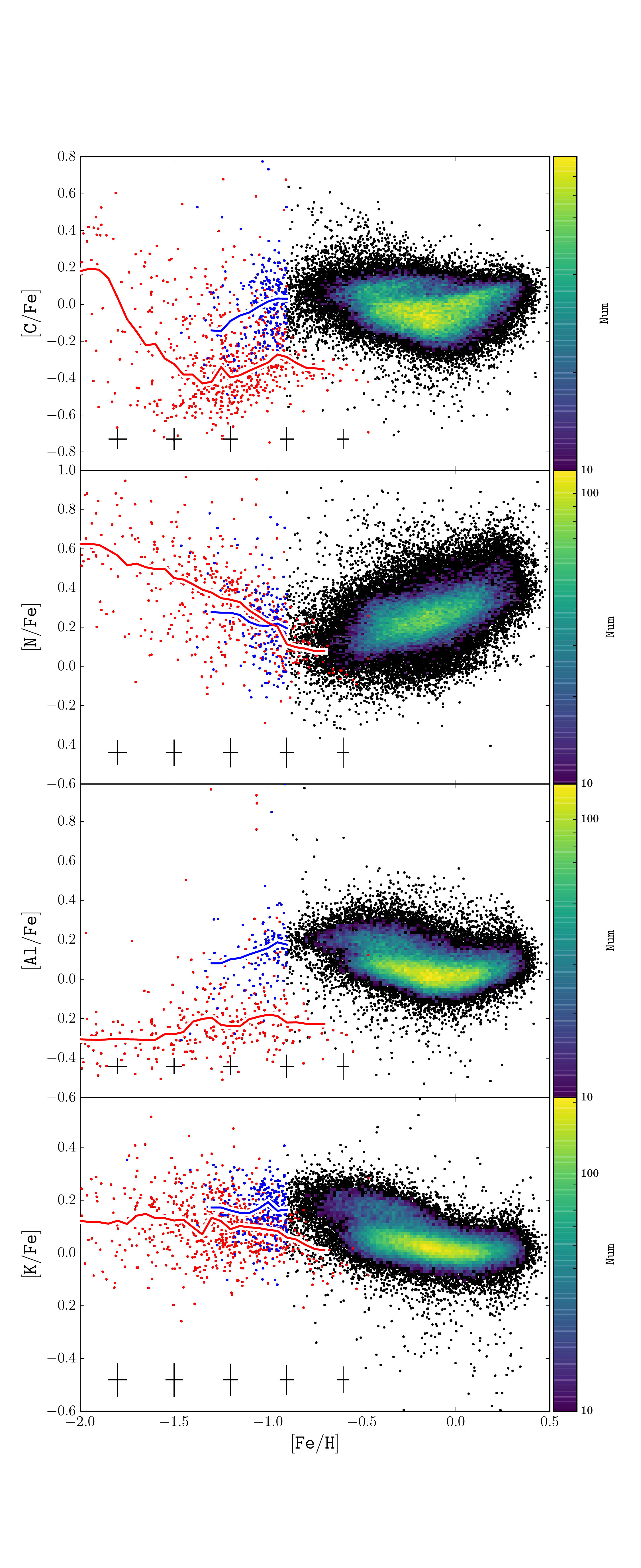}
  \caption{Same as Figure \ref{alphagold}, but for the abundance distributions of the light and odd-Z elements C, N, Al, and K.  The element here for which the two metal-poor populations stand out most distinctly from one another is Al, where the LMg population appears Al-poor and the HMg population has approximately solar Al levels.}
  \label{elemplot1}
\end{figure}

\begin{figure}
  \centering
  \includegraphics[scale=0.4,trim = 0mm 20mm 0mm 8mm, clip]{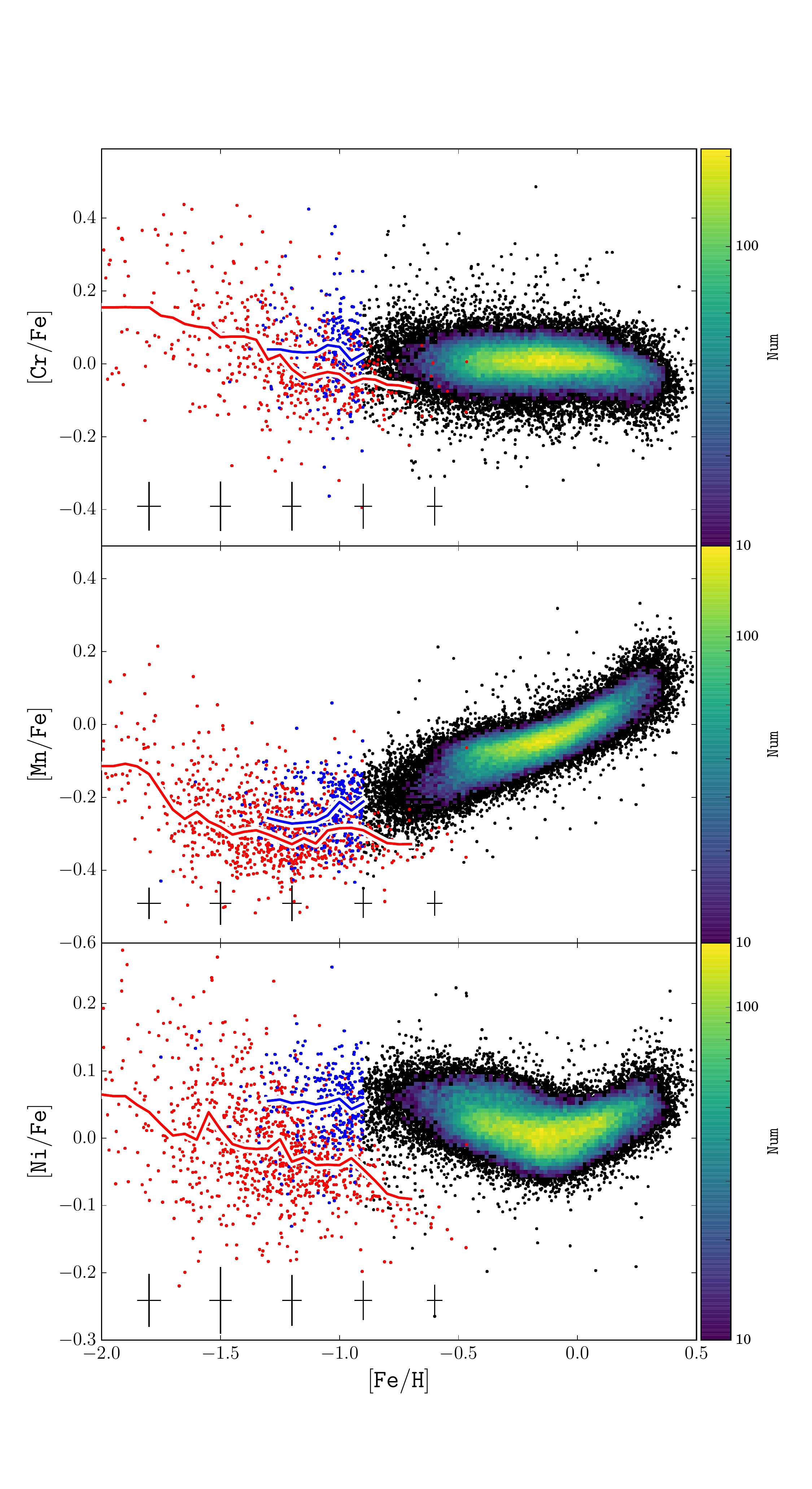}
  \caption{Same as Figure \ref{alphagold}, but for the abundance distributions of the iron peak elements Cr, Mn, and Ni.  Here we see some significant separation of the LMg and HMg populations in Ni, in a fashion similar to that seen in some of the  $\alpha$-elements.}
  \label{elemplot2}
\end{figure}

\subsubsection{Light and Odd-Z elements:  C, N, Al, and K}

Figure \ref{elemplot1} presents some of the light and odd-Z element patterns, and demonstrates a very distinct separation in aluminum at the level of almost $\sim 0.5$ dex.  The population with lower Mg (the LMg population) is found to be Al-poor, and the higher Mg population (the HMg population) has solar-level to slightly above solar enriched levels of Al, which is consistent with the results from the smaller more metal-rich sample from \citet{haw15}.  The gap in Al between the two populations is remarkably large.  While it is tempting to use Al as the primary discriminating element for low metallicity populations, we refrain from doing so now for two reasons (1) the typically larger ASPCAP uncertainties on [Al/Fe] ratios, and (2) much smaller sample sizes when selecting stars based on Al abundances rather than Mg abundances because fewer stars have the low uncertainties necessary to make fine chemical distinctions.  However, future studies with better aluminum data may find great power in using this element as a discriminator of these two metal-poor populations.

While there is significantly more scatter, we also see some distinction between these two populations in carbon, with the LMg stars typically having lower [C/Fe] (for a given metallicity), although first dredge-up and subsequent mixing for stars at the red giant branch (RGB) bump will bring up CNO-cycle processed material, typically decreasing the surface [C/N] from its natal level through a decrease in $^{12}$C and an increase in $^{14}$N \citep{gra00,kl14}.  In the metallicity range where our two populations overlap, we find that their typical [N/Fe] ratios are quite similar.  At lower metallicities the N-abundances increase in the LMg stars, however, because these are measured from CN features in APOGEE spectra, which are weak and can disappear for very metal-poor stars \citep{hol15}, these [N/Fe] ratios are very uncertain (and instead many are upper limits).  Note that unlike other chemical abundances from APOGEE, which are calibrated to remove abundance trends with temperature seen in cluster stars \citep{hol15}, C and N are not calibrated in this way because dredge-up and mixing in giants intrinsically produces trends with temperature. The data exhibit only a slight difference in the median [K/Fe] ratios between the LMg and HMg populations, which is diluted by the large scatter in both populations, such that their [K/Fe] distributions appear similar.

\subsubsection{Iron Peak Elements:  Cr, Mn, and Ni}

Figure \ref{elemplot2} presents the distributions of heavier, iron-peak elements.  The most significant separation between these two populations is in Ni, which, although less pronounced than for Al, is similar in appearance to some of the $\alpha$-element abundance separations (as shown in Figure \ref{alphagold}).  For Ni we see a slight overlap of the two populations, but most HMg stars have higher Ni abundances than the LMg stars, again consistent with the findings of \citet{haw15}.  Like K, the distributions of Cr and Mn abundances for the two populations mostly overlap, although the median of these distributions show slight differences, with the LMg population having lower [Cr/Fe] and [Mn/Fe] ratios.  The Cr and Mn distributions appear to be flat or decreasing with decreasing metallicity above \feh \ $\gtrsim -1.4$, but at lower metallicities the LMg population appears to begin increasing in Cr and Mn with decreasing metallicity.

For Cr and Mn, as well as for Ni and possibly C, at low metallicities there is a slight increase in [X/Fe] ratios with decreasing metallicity.  The likely reason for this trend is that, at low metallicities, the lines used to measure these elements become increasingly weaker, so much so that they should be undetectable at the typical S/N of our selection criteria.  Thus, the measurements presented at the lowest metallicities would instead be upper limits.  We would then expect to see an increasing [X/Fe] trend with decreasing metallicity because of two effects.  (1) When measuring upper limits, ASPCAP is effectively fitting the noise present in spectra, so for a set of stars with a range in temperature, we expect the derived upper limits to be temperature dependent.  Hotter stars have intrinsically weaker lines, so when fitting the same noise level, higher abundances will be derived for these stars than for cool stars, which should have stronger lines for a given abundance.  This has the effect of artificially increasing the median abundance ratio at low metallicities where these upper limits appear.  (2) We also expect lower abundance ratios to have higher reported uncertainties, and thus more likely to result in a star being cut out at these low metallicities by our maximal uncertainty criterion.  By tending to remove stars with lower and less certain abundances, we artificially drive up the median abundance ratios at the lowest metallicities (as we approach [Fe/H] $\sim -2.0$), and we expect this to affect especially elements like Cr, Mn, Ni and possibly C.

These effects should primarily impact Cr, which has the weakest lines of the four elements mentioned above, the abundances reported are more likely to represent upper limits below metallicities of around [Fe/H] $\sim -1.5$ or $-1.6$.  Cr would then be followed by Mn and Ni, which would both return upper limits at even lower metallicities.  On the other hand, for carbon it would be more surprising if overestimated line strengths are responsible for the up turn at low metallicities, because there are so many carbon features throughout the APOGEE spectra that are used to derive [C/Fe].   More of the [C/Fe] ratio measurements, therefore, may be real even at lower metallicities.  Moreover, stars with high [C/Fe] ratios are not unexpected given the existence of carbon-enhanced metal-poor (CEMP) stars with [C/Fe] $> +1.0$ \citep[see][and references therein]{bc05,fn15}, so the trends seen in [C/Fe] may well be real.

\subsubsection{Combined Light Elements:  (C+N) and (C+N+O)}

\begin{figure*}
  \centering
  \includegraphics[scale=0.4]{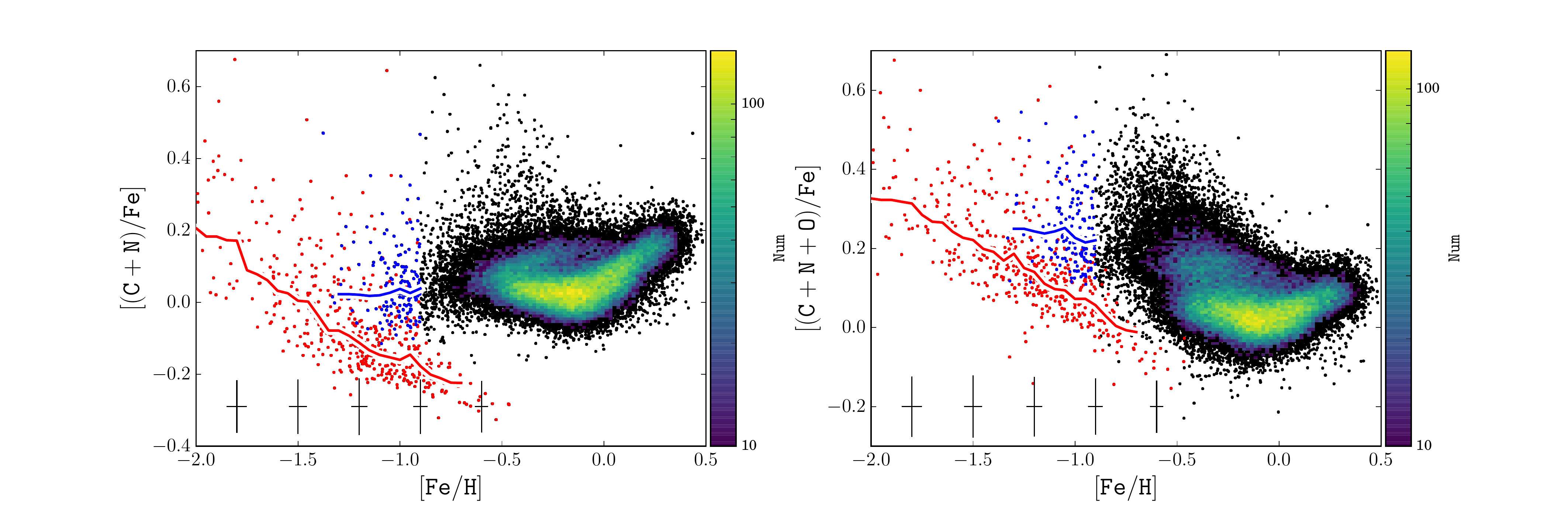}
  \caption{Same as Figure \ref{alphagold}, but for the combinations of C, N, and O.  There is a relatively high degree of separation of the LMg and HMg populations in both C+N and C+N+O, as well as distinct trends within each population, i.e., the decreasing [(C+N)/Fe] and [(C+N+O)/Fe] ratios with increasing metallicity in the LMg population, and nearly constant ratios in the HMg population.}
  \label{cncno}
\end{figure*}

As mentioned above, first dredge-up and mixing at the RGB bump can effect the surface abundances of (primarily) carbon and nitrogen \citep[with small changes possible for oxygen;][]{gra00,kl14,mar16}, so that these abundances no longer reflect their natal values.  However, the total abundance of carbon, nitrogen, and oxygen (as represented by [(C+N+O)/Fe]) should remain unchanged by the dredge-up, and since the birth oxygen abundance is nearly conserved in low-mass stars, the surface [(C+N)/Fe] is also essentially unchanged \citep{gra00,mar16}.  Figure \ref{cncno} shows that most of the stars in the LMg and HMg populations exhibit different C+N and C+N+O abundances, and they are quite distinct.  In both of these abundances, the HMg exhibits a scatter around solar [(C+N)/Fe] and [(C+N+O)/Fe] of $+0.2$ dex.  For the same metallicity, the LMg shows lower C+N and C+N+O abundances than the HMg, which provides another example of a chemical space where the LMg and HMg populations appear to separate reasonably.  

The LMg stars exhibit decreasing C+N and C+N+O abundances with increasing metallicity in addition to having a decreasing scatter with increasing metallicity.  The higher scatter at lower metallicities is most likely due to less reliable abundance measurements from weaker lines (primarily poor N abundances from weak CN lines) in metal-poor stars, but the concentration in LMg abundance ratios especially narrows for [Fe/H] $ \gtrsim -1.3$.  This tight trend then continues to metallicities higher than our initial examination cutoff at [Fe/H] $ = -0.9$, and we can see a (C+N)-poor group of ``LMg-extension'' stars that reaches to [Fe/H] $\sim -0.5$.  These stars appear to follow the chemical abundance pattern set by the metal-poor LMg stars, but have [(C+N)/Fe] (and to some extent even [(C+N+O)/Fe]) ratios that deviate significantly from the canonical thin and thick disk populations, which have [(C+N)/Fe] ratios nearly at or greater than solar.  Therefore, we assign these stars to the LMg population.  We do so explicitly by examining the [Fe/H] $> -0.9$ stars with subsolar [(C+N)/Fe] and assigning a conservative, by-eye linear expression for the upper [(C+N)/Fe] envelope of this distribution (given numerically by [(C+N)/Fe] $= -0.4$ [Fe/H] $- 0.46$).  We then define stars with [(C+N)/Fe] under this envelope as potential LMg stars, but note that this is a conservative selection, again to avoid thin disk contamination.  These more metal-rich LMg stars are also shown in previous figures, where they appear to follow other LMg population trends, and support that these stars are members of this population.

\subsection{Exploring Multi-Dimensional Chemical Space}

\begin{figure}
  \centering
  \includegraphics[scale=0.4]{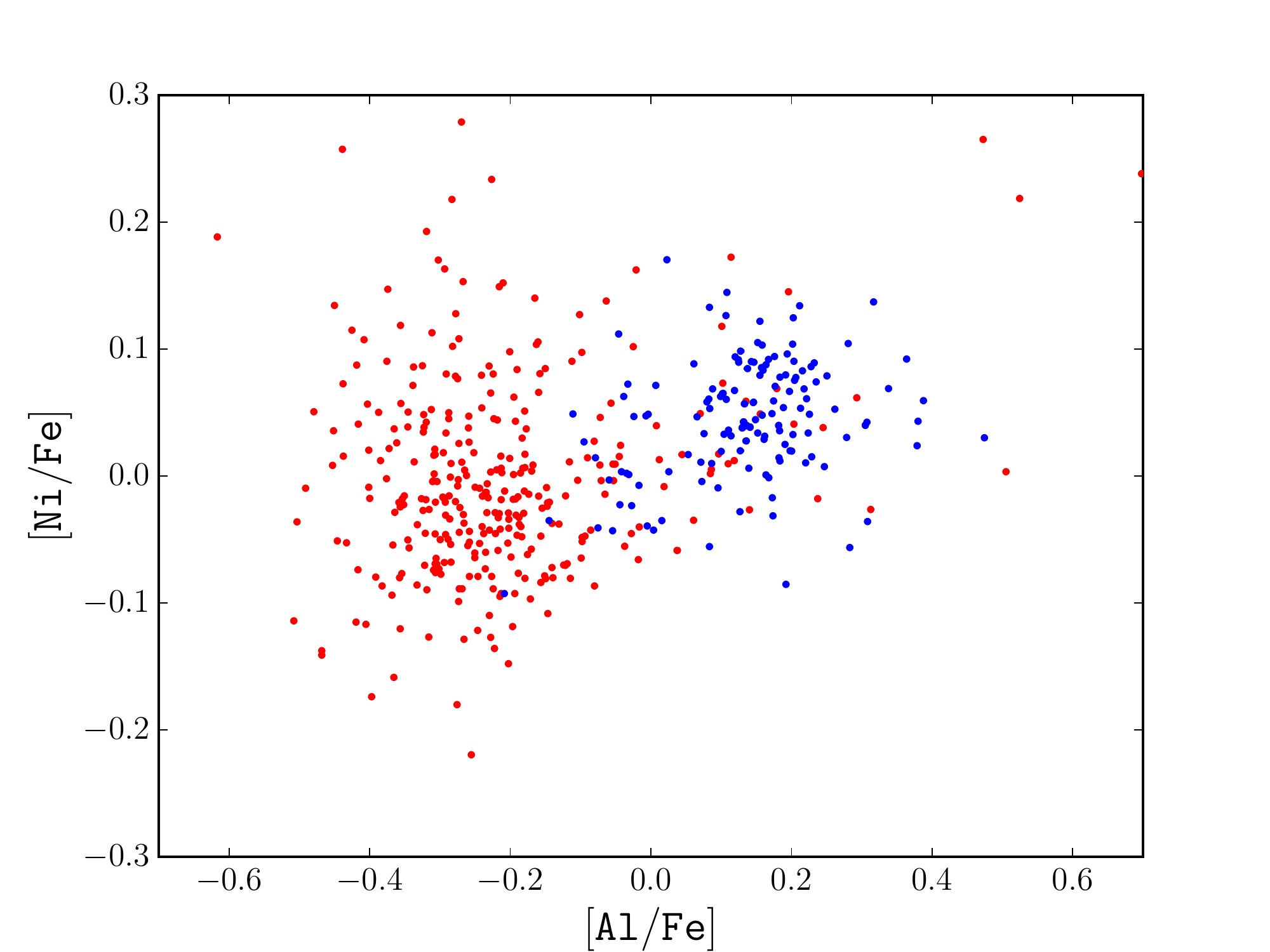}
  \caption{ [Ni/Fe] vs. [Al/Fe] for metal-poor stars ([Fe/H] $< -0.9$) in the LMg (red) and HMg (blue) populations shown in Figure \ref{mgfegold}.  This slice of chemical space is one example where these two populations cluster with good separation, and demonstrate how incorporating different chemical information provides opportunities for further refining the definitions of populations.}
  \label{NiFeAlFe}
\end{figure}

\begin{figure}
  \centering
  \includegraphics[scale=0.4]{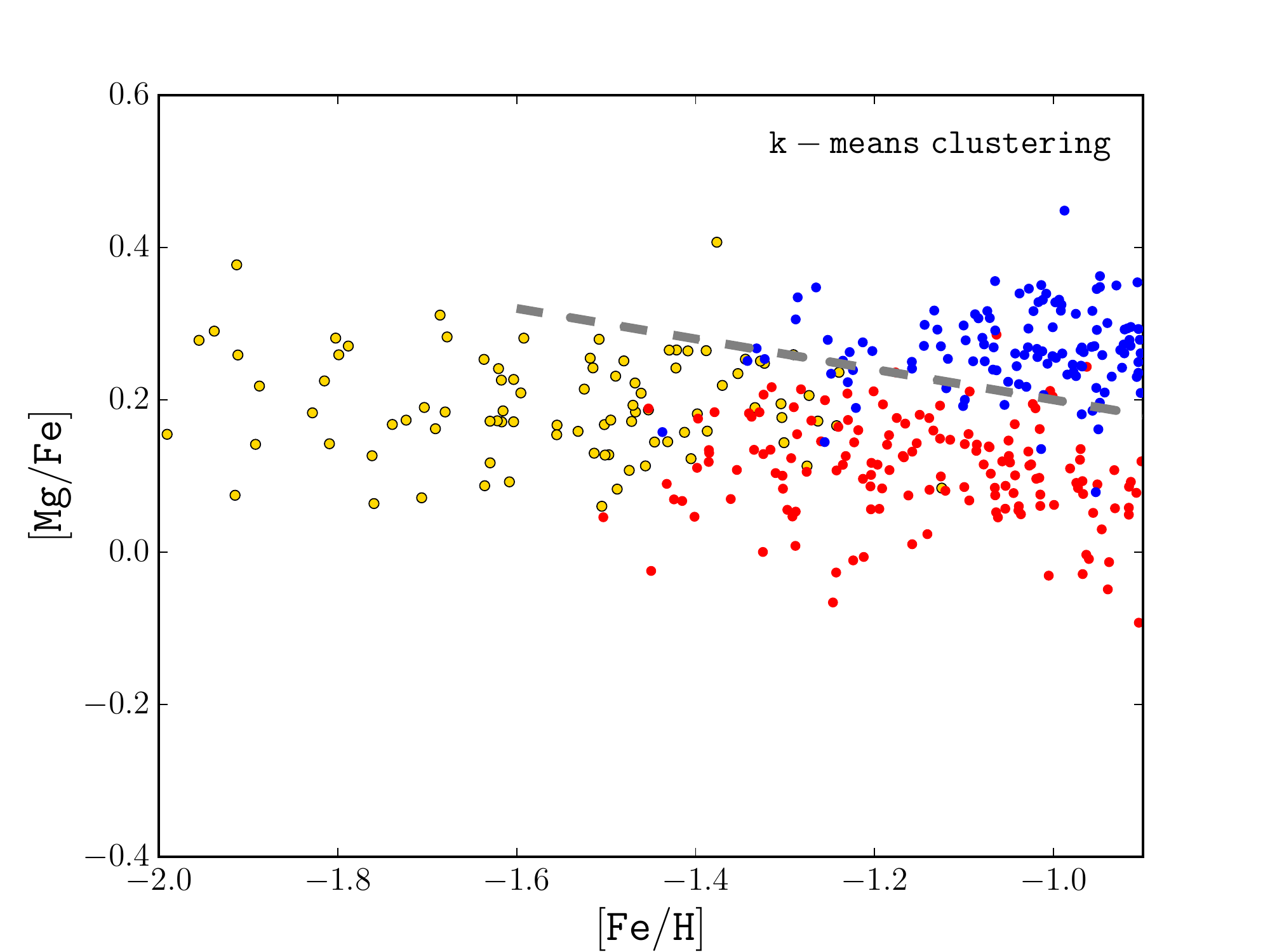}
  \caption{ [Mg/Fe] vs. [Fe/H] projection of the 11-D chemical space probed, where we have performed clustering analyses on stars with well defined chemical abundances, as described in the text.  Stars are color-coded by cluster assignment according to the k-means clustering algorithm, two of which (colored red and blue) are very similar to the two populations that we defined by eye in Figure \ref{mgfegold}.}
  \label{clustering}
\end{figure}


As demonstrated, the two metal-poor populations we have identified via their [Mg/Fe] distributions in the APOGEE database are also quite well discriminated in other elemental ratios, such as [Al/Fe] and [(C+N)/Fe].  While we have examined stellar abundances of different elements one by one, these stars live in a highly multi-dimensional chemical space that can be sliced in many different ways to search for distinct stellar populations.  For example, the [Ni/Fe] versus [Al/Fe]-plane for stars attributed to the LMg and HMg populations in Figure \ref{NiFeAlFe}, shows a striking separation.  This is similar to the separation reported by \citet{ns10}, who examined [Ni/Fe] versus [Na/Fe] for stars between [Fe/H] $ = -1.6$ and $-0.4$, and found two populations (which they also selected based on the [Mg/Fe]-[Fe/H]) like those examined here.  This perhaps should be expected, because Na and Al are produced through the NeNa and MgAl cycles, which are linked and operate under similar temperature ranges \citep{arn99}; thus Na and Al abundances should be roughly correlated.  

But again, Figure \ref{NiFeAlFe}, like all preceding figures are just two-dimensional slices through chemical space, when we have many more dimensions that we can utilize simultaneously.  While it is difficult to visualize higher dimensionalities, we can use tools such as clustering algorithms to search this space to provide statistically rigorous tests of our proposed separations.

To conduct such a multidimensional probe and to quantify how well the two populations and their differences are captured by our simple selection in [Mg/Fe] vs. [Fe/H], we utilize two clustering algorithms to independently and objectively look for these populations.  The multi-dimensional space we search is that of metallicity ([Fe/H]), [(C+N)/Fe] (which should be more representative of birth abundances than C or N separately due to the effects of first dredge-up, as discussed earlier), and [X/Fe] for O, Mg, Al, Si, K, Ca, Cr, Mn, and Ni, i.e., those elements with good data that were previously examined.  

First, we use an algorithm to perform k-means clustering \citep{mac67} to search for clusters in the aforementioned 11-dimensional chemical space for all stars with uncertainties under 0.1 dex and [Fe/H] $ \leq -0.9$ (so that the populations noted in this work are not lost to the much more populous thin and thick disk chemical distributions).  We performed a silhouette analysis \citep{rou87} to determine the optimal number of clusters ($k$) to represent the data, finding that three clusters best describe the data.  The resulting assignment of stars for the three clusters are shown in the [Mg/Fe] - [Fe/H] plane in the left panel of Figure \ref{clustering} color-coded according to their cluster assignment by the k-means algorithm.  

Of the three clusters identified, two seem to separate primarily in metallicity from the third cluster of stars typically having [Fe/H] $\lesssim -1.4$ and entirely representing the lowest metallicity stars.  This third cluster may reflect the fact that at metallicities below [Fe/H] $\sim -1.5$ the LMg and HMg populations blend together into one chemically indistinguishable group.  Alternatively this third group may be a spurious bifurcation of one of the other two clusters (presumably the cluster corresponding to the LMg), either as an artifact of the k-means algorithm or due to low statistics creating a small gap in an otherwise continuous sequence.  Whether there may be an astrophysical reason for this {\it distinct}, metal poor population should be reconsidered if it persists despite more data or improved techniques applied to this problem.

The remaining two k-means clusters are located at higher metallicities, where the LMg and HMg are more distinct.  How these clusters relate to the populations defined by our visual inspection of only the two-dimensional [Mg/Fe]-[Fe/H] plane is shown by the over-plotted dividing line we initially used to separate the LMg and HMg populations in Figures \ref{contour} and \ref{mgfegold}.  As may be seen, our adopted dividing line appears to properly separate most of the stars assigned to either of the more metal-rich clusters defined independently by the k-means algorithm.  

Apart from this metal-poor cluster, we note that the k-means clustering has produced one relatively low- and one relatively high-Mg clusters.  Specifically, we find that of the stars below and above the line in Figure \ref{clustering} respectively, 90\% (146/163) of the LMg population stars are assigned to the low-Mg k-means cluster and 95\% (103/108) of the HMg stars are assigned to the high-Mg k-means cluster.  Thus, the k-means algorithm identifies clusters relatively consistent (at least at the higher metallicity range of our sample) with the two populations we specified using our by-eye division based on only two chemical dimensions.  This suggests that [Mg/Fe] and metallicity alone are a robust discriminator of the two groups of relatively metal-poor stars.  We also note that most of the cross-contamination occurs around our dividing line, and somewhat at the metal-poor end of the high-Mg or HMg population distribution, where the third, metal-poor cluster dominates.

The other clustering algorithm that we try is DBSCAN \citep{est96}, a two parameter density-based clustering algorithm that builds clusters by chaining together data points that have a minimum of $N$ neighbors within a multi-dimensional sphere of radius $\epsilon$.  Together, these parameters determine a minimum ``density'' and the algorithm identifies clusters present in the data with that density.  Because this algorithm is density based, it will tend to exclude data on the outskirts of clusters, and can be  fairly sensitive to the choice of input parameters (which effectively define the desired densities of output clusters).  Nevertheless, DBSCAN also delivers results consistent with our by-eye selection in finding two clusters (with input parameters of $N = 17$ and $\epsilon = 0.21$ dex).


Of the stars assigned to the two DBSCAN clusters 94\% (119/127) of the low Mg abundance cluster stars would be properly associated with the LMg population according to the by-eye definition, and 91\% (64/70) of the stars assigned to the high Mg abundance cluster would be identified as HMg population stars.  The remaining 163 stars lie in less densely populated regions of chemical space than the cores of the clusters and are thus unassigned to either of these clusters.   Because of this, as was the case with the k-means clustering analysis, the two clusters found by DBSCAN that seem to correspond to the LMg and HMg populations are predominately populated at the higher metallicities of this sample ([Fe/H] $\gtrsim -1.5$), leaving the lower metallicity stars unassigned.  While our adopted values of the DBSCAN N and $\epsilon$ parameters are not definitive (and indeed alternative pairings produce similar clusters as those shown in Figure \ref{clustering}), DBSCAN clustering analysis reveals that there is a density threshold that produces two distinct clusters with a manner of separation that is consistent with our initial separation in the [Mg/Fe]-[Fe/H] plane.


In summary, we find that the k-means and DBSCAN algorithms reaffirm our by-eye discrimination, and identify very similar clusters to those we identify as the LMg and HMg populations at the metallicities where we see the largest differences in chemical distributions.  This is an objective affirmation that these populations are real, and that our method to separate them in a single projection of the [Mg/Fe]-[Fe/H] plane properly assigns 90\% or more of stars to the correct population as compared to the results of clustering algorithms.  While for simplicity we proceed with the use of a strict two-dimensional, Mg-based division of the two metal-poor populations, there will naturally be a small degree of cross-contamination (as we saw with the comparison to the k-means and DBSCAN results), due to some intrinsic overlap of these populations, the projection of a multi-dimensional distribution into two dimensions, and uncertainties blurring the intrinsic distribution of these populations.  In the future, when truly large samples of multi-dimensional data are available for metal-poor stars, purer discrimination will be possible by looking at multiple chemical dimensions.

\subsection{Kinematical Nature of the LMg and HMg Populations}
		
\begin{figure*}
  \centering
  \includegraphics[scale=0.5]{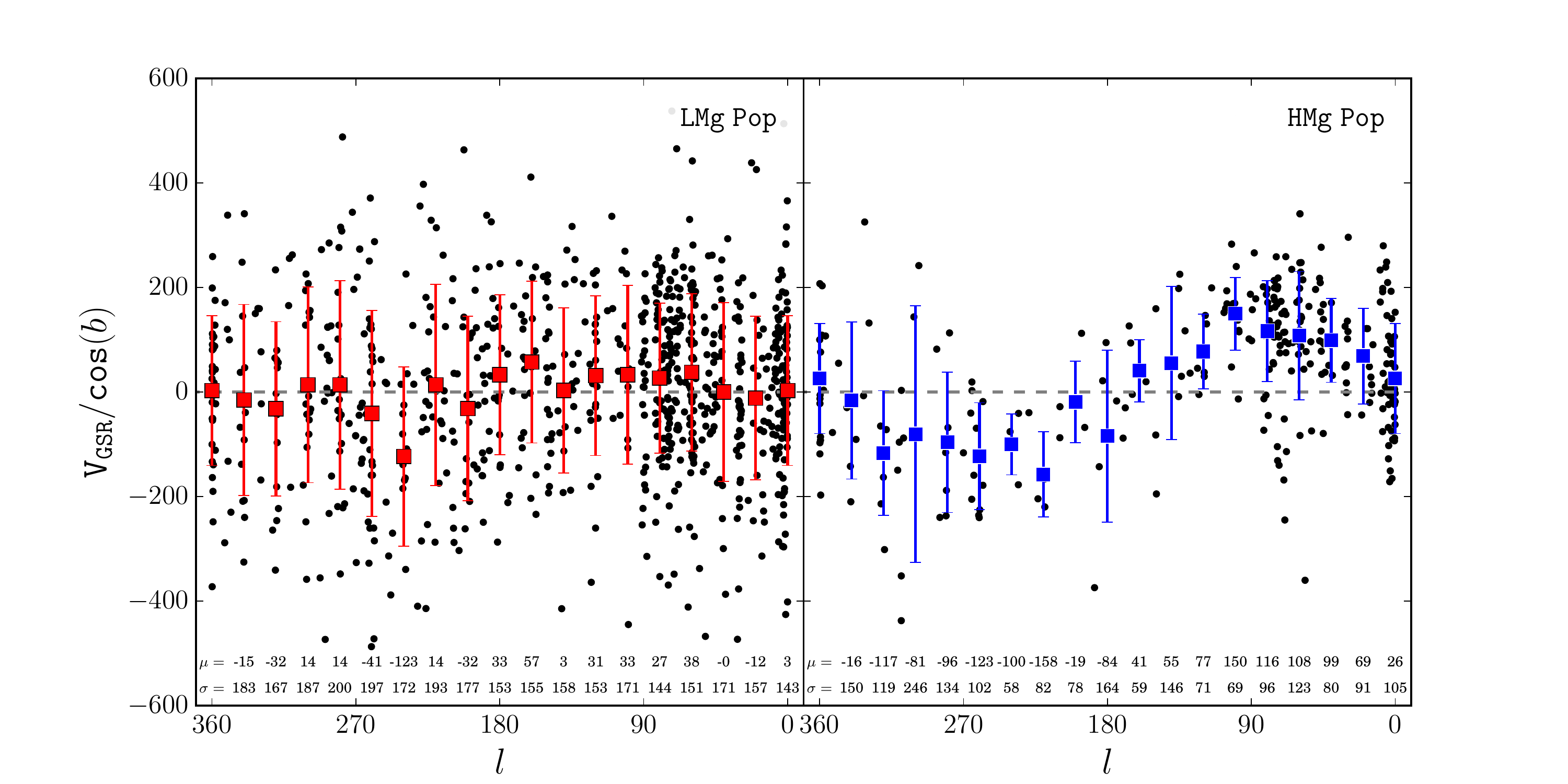}
  \caption{$V_{GSR}/\cos(b)$ vs. Galactic longitude for the LMg population (left) and the HMg population (right).  The colored symbols represent the mean and population standard deviation calculated for 20\degree \ bins (the $l = 0$\degree \ $ = 360$\degree\  bin is repeated on either end), after applying a $3\sigma$ cut to remove stars with potentially errant velocities.  The means and standard deviations of each bin are shown at the center of the bin at the bottom of the plot.  Here we can see that the LMg population has on overall halo-like distribution of velocities, i.e. a large dispersion with little to no systemic rotation.  The HMg population, on the other hand, appears to have a significant rotation with a much smaller dispersion. }
  \label{vgsrcosb}
\end{figure*}

While these two populations appear chemically distinct, they would be even more astronomically significant if they additionally exhibit different kinematics, which we can examine using the radial velocities of stars measured by APOGEE.  We convert these radial velocity into the Galactic Standard of Rest system assuming a solar motion of $(V_r , V_{\phi}, V_z)_\odot = (14, 250, 7)$ \kms \ \citep{sch10,scho12}.  \citet{maj12} have shown the utility of the Galactic longitude-$V_{\rm GSR}/\cos(b)$ diagram for revealing stellar populations kinematically using only radial velocity data.  $V_{\rm GSR}/\cos(b)$ is a proxy for the planar velocity of a star projected onto our line of sight, but breaks down at high Galactic latitudes \citep{maj12}, so in this examination we only use stars with $|b| < 62$\degree \ (to include the stars in the APOGEE fields centered at $b = 60$\degree).

Figure \ref{vgsrcosb} shows that the distribution of $V_{\rm GSR}/\cos(b)$ vs. Galactic longitude for the LMg population has a large velocity dispersion (roughly 150-200 \kms, drawn from Figure \ref{vgsrcosb}) with very little to no net rotation, typical of that expected for a halo population.  The HMg population, on the other hand, has a modest velocity dispersion of about 80-120 \kms \ around a significant trend of net rotation at the level of about 120-150 \kms \ (taken from the amplitude of the sinusoidal velocity variation displayed in ther right panel of Figure \ref{vgsrcosb}); the latter is consistent with the rotational velocity for the thick disk, at least at lower metallicities \citep{cb00,lee11,ad13,ap16}.  This is perhaps unsurprising, since chemically, the HMg population looks like an extension of the thick disk.  Included in the HMg population are a few stars that have radial velocities more typical of halo-like kinematics, which may be halo stars with chemistry similar to the thick disk, or contamination from the LMg population.

\begin{figure*}
  \centering
  \includegraphics[scale=0.5]{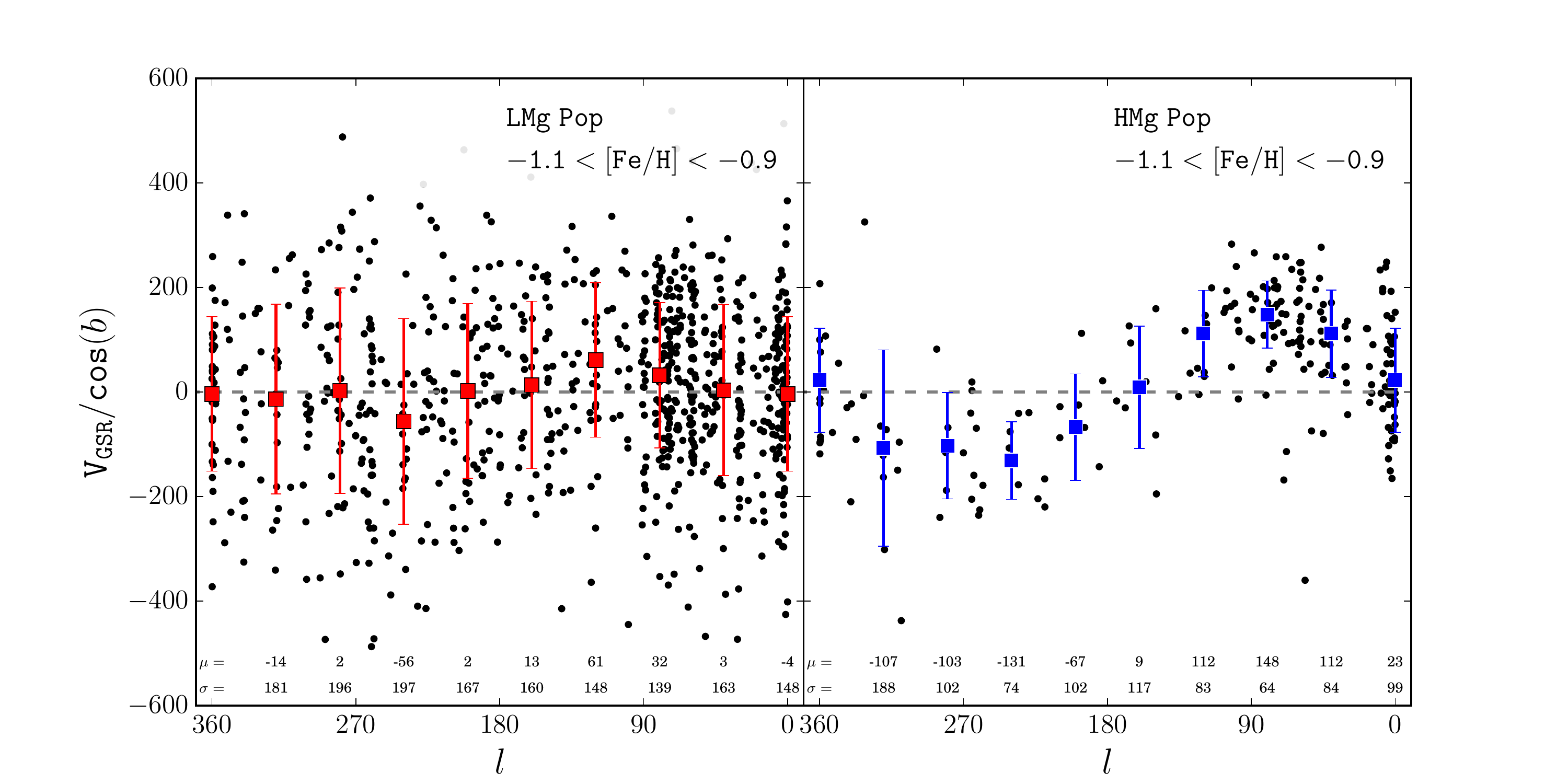}
  \caption{Same as Figure \ref{vgsrcosb}, with data binned into 40\degree \ bins, for LMg and HMg stars with $-1.1 <$ [Fe/H] $<-0.9$.  The more metal-rich ends of these populations follow the same kinematical trends as the subsamples covering the larger metallicity range shown in Figure \ref{vgsrcosb}.}
  \label{vgsrcosb_metalsep}
\end{figure*}

Because the LMg population spans a wider range in metallicity than the HMg population and one would expect more stellar contribution from the halo (rather than the disk) towards lower metallicities, it is of interest to confirm that the above kinematical signatures persist even at the higher metallicity end of our samples.  To do so, we examine $V_{GSR}/\cos(b)$ vs. Galactic longitude only for stars with metallicities [Fe/H] $ > -1.1$ in each of these populations (Figure \ref{vgsrcosb_metalsep}).  Acknowledging the much smaller net samples, we still find that even the more metal-rich stars of the LMg exhibit halo-like motions, which further justifies that the LMg population is a coherent and distinct population from the dynamically colder HMg population.
		
\section{Discussion}

\subsection{Relation to High-$\alpha$ and Low-$\alpha$ Halo Stars}

\begin{figure*}
  \centering
  \includegraphics[scale=0.4,trim = 0mm 20mm 0mm 8mm, clip]{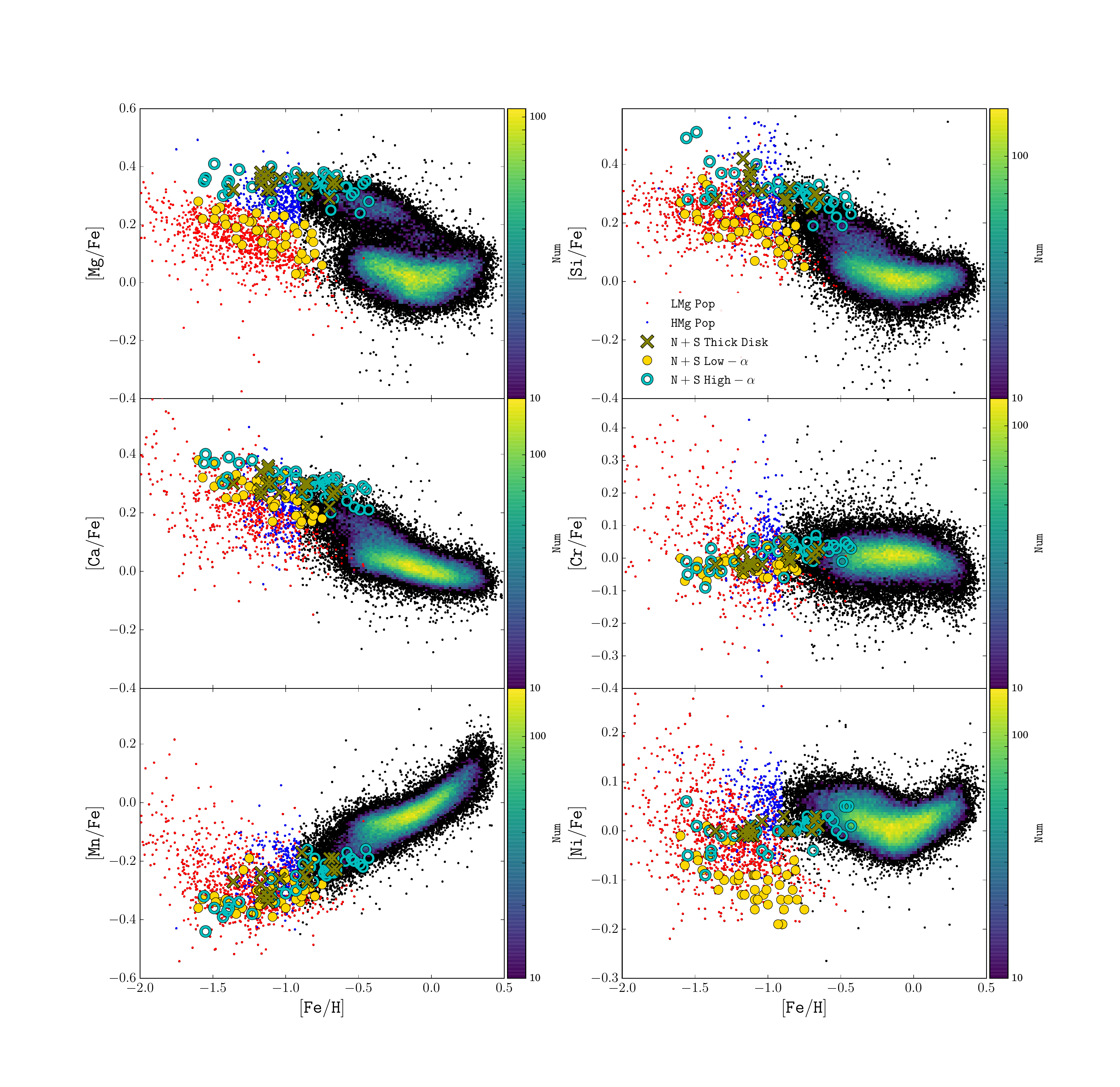}
  \caption{Distribution of [X/Fe] with [Fe/H] for Mg, Si, Ca, Cr, Mn, and Ni with 2D histogram of the densely distributed stars as done in Figure \ref{mgfe}.  Stars of LMg and HMg populations are color-coded the same as in Figure \ref{mgfegold}.  Over-plotted are data from \citet{ns10,ns11} color-coded to identify kinematically selected thick disk stars (olive green crosses), and their chemically selected high-$\alpha$ (cyan open circles) and low-$\alpha$ (yellow filled circles) halo stars.}
  \label{ns}
\end{figure*}

Our analysis of the APOGEE database has focused on a very specific examination of a large sample of metal-poor stars making use of the clear separation seen in the [Mg/Fe]-[Fe/H] plane, a separation also validated in other chemical dimensions, like [Al/Fe]-[Fe/H] and [(C+N)/Fe]-[Fe/H], as well as in the overall 11-D chemical space (see Section 3.3).  Previous studies of smaller samples of stars have demonstrated a split of halo stars into high- and low-$\alpha$ groups \citep{ns10,ns11,nav11,ram12,sch12,she12,jj14,haw15}.  These groups appear generally to correspond well with our HMg and LMg populations, as we now demonstrate.

In a study of the abundances of $\alpha$-elements (Mg, Si, Ca, Ti), Na, Cr, and Ni for 94 kinematically and metallicity selected dwarf stars, \citet{ns10} found two populations of stars with halo-like kinematics (total space velocities, $V_{\rm tot} > 180$ \kms) separated in the \mgfe-\feh \ plane.  The population of halo stars with lower \mgfe \ also separated from the higher \mgfe \ ratio population in other $\alpha$-elements \citep[although to a lesser extent;][]{ns10,ram12,haw15}, and other elements, such as (C+N), Na, Al, Ni, Cu, Zn, Y, and Ba, whereas little to no distinction was seen for other elements such as Cr and Mn \citep{ns10,ns11,haw15}. 

Figure \ref{ns} compares the chemistry of the APOGEE DR13 sample of this study to the stellar abundances presented in \citet{ns10,ns11}.  We find general agreement for most elements, with perhaps small offsets between the two data sets in a few cases.  The largest differences may be seen in the distribution of [Ca/Fe] ratios of the two populations seen here, which is likely due to the different methods of spectroscopic analysis employed by \citet{ns10,ns11} and APOGEE.  \citet{ns10,ns11} measured abundances relative to two bright thick disk stars to achieve a high internal precision, but may be subject to systematic offsets compared to chemical abundance measurements using different methods, such as APOGEE's automated spectroscopic analysis.  This difference in method of analysis, along with the use of differing spectral lines, model atmospheres, etc. may lead to the offsets seen in [Ca/Fe] ratios as well as those in other elements.

In addition to the chemical similarities between the low- and high-$\alpha$ halo stars and our HMg and LMg populations, there are kinematical similarities linking these groups of stars.  Both the thick disk and high-$\alpha$ halo stars have (on average) higher rotational velocities than the low-$\alpha$ halo stars \citep{ns10}, analogous to the kinematical differences seen between the HMg and the LMg populations here (see Figure \ref{vgsrcosb}).  Thus, kinematics affirm that the low-$\alpha$ halo stars are members of the same population as the LMg population stars identified here, and that the high-$\alpha$ halo stars are part of the HMg population.  While the LMg/low-$\alpha$ halo stars and HMg/high-$\alpha$ halo stars seem to be samples of the same respective populations, we maintain the usage of the names LMg and HMg to more explicitly reflect their selection through Mg abundances, the $\alpha$-element that most easily distinguishes these populations.

It is interesting that there is such good agreement between the samples of stars in our work and  \citet{ns10,ns11}, given the vastly different volumes sampled by each work.  APOGEE surveys a large volume, allowing it to reach into the bulge or out into the halo.  In contrast, \citet{ns10,ns11} studied a sample of stars from the solar neighborhood, extending only as far as $\sim$ 335 pc.  The fact that both studies find similar distributions of stars suggests that they come from parent populations that, in terms of their distribution, do not vary significantly with position in the Galaxy.  By concluding that the populations studied here are the same or related to those revealed by the high- and low-$\alpha$ halo stars \citep[initially seen by][]{ns10}, APOGEE uses its large statistical sampling to bring more clarity and significance to these two distinct populations.  

\subsection{Comparison to Milky Way Satellites}


\begin{figure*}
  \centering
  \includegraphics[scale=0.4,trim = 0mm 20mm 0mm 30mm, clip]{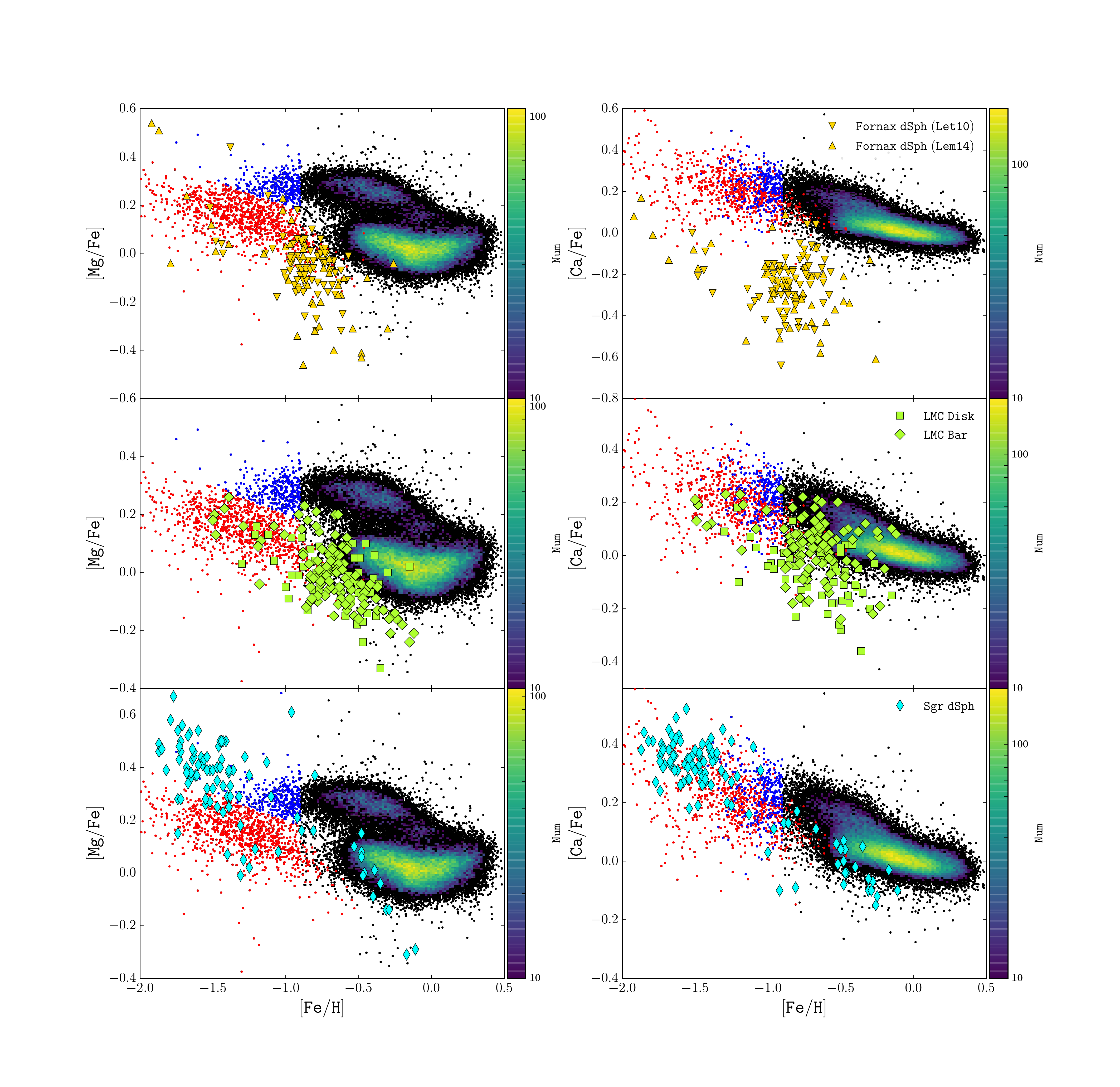}
  \caption{Distribution of [X/Fe] with metallicity for Mg (left) and Ca (right).  Stars of the LMg and HMg populations are color-coded the same as in Figure \ref{mgfegold}.  Over-plotted are abundances of (Top) Fornax dSph stars reported by \citet[][yellow downward triangles]{let10} and \citet[][yellow upward triangles]{lem14}, (Middle) LMC disk (green squares) and bar (green wide diamonds) stars from \citet{vds13}, and (Bottom) Sgr dSph and M54 stars (cyan narrow diamonds) reported by \citet{muc17}.}
  \label{dsph}
\end{figure*}




One possible origin for metal-poor stars in the MW is through the accretion of smaller, dwarf spheroidal (dSph) systems.  As noted in the past \citep{venn04,tol09} dSph stars typically have lower $\alpha$-element abundances than most MW stars at the same metallicities.  However, at lower metallicities ([Fe/H] $\lesssim -1.5$) there is more overlap in [X/Fe] between the chemistry of MW and dSph stars.  This suggests that, at least at higher metallicities, dSph stars from lower mass dwarf galaxies like those common around the Milky Way, are unlikely to contribute significantly to either the LMg or HMg populations.  This does not, however, rule out the possibility that satellite galaxies could have contributed stars to our halo with different chemistry or that dSphs could have contributed stars at lower metallicities where the agreement is better.  

The sample of dSph stars examined in some of these past studies come from multiple dSph galaxies.  While this gives us an idea of the general spread of abundances across dSph satellites, it does not provide a picture of the chemical evolution within a given satellite.  If we want to assess the dSph populations that are more likely to have been contributed to the MW, we should focus on the chemical evolution in more massive satellites, because previous studies \citep{bujo05,font06} have found that satellites accreted earlier in the history of a galaxy are expected to be, on average, more massive.  This is likely to have an impact on the chemistry of these satellites, because we might expect more massive satellites to have experienced more enrichment before Type Ia supernovae began to contribute their yields to the interstellar medium (e.g., due to higher star formation rates, higher star formation efficiency, better retention of supernovae products, etc.).  This would have the effect of pushing the \alphafe-knee of these satellites to higher metallicities leading to potentially higher \alphafe \ ratios than less massive satellites for a given metallicity, and resulting in better agreement with the metal-poor stars seen in the MW at a given metallicity.  

Although a few of these more massive satellites were somewhat represented in past studies, we wish to compare the APOGEE sample to a larger set of abundances from one of them, --- the Fornax dSph --- by examining the red giant abundances measured from high-resolution spectra by \citet{let10} and \citet{lem14}.  We also compare our APOGEE sample to the chemical abundances of Large Magellanic Cloud (LMC) red giants derived from high-resolution spectra by \citet{vds13} and of Sagittarius (Sgr) dSph and M54 stars measured from medium-resolution spectra by \citet{muc17}.  We show the of Mg and Ca abundance distributions for each of these systems in comparison to our APOGEE sample in Figure \ref{dsph}.  These two chemical elements show trends where chemical abundance pattern differences appear to show up most distinguished either amongst Milky Way populations or between satellites and the Milky Way.

The top panels of Figure \ref{dsph} show that Fornax stars exhibit [Mg/Fe] ratios on the low side of the LMg population's chemical abundance pattern, except at the lowest metallicities where the agreement is better. On the other hand, the Fornax Ca abundances do not agree well with most of the stars observed by APOGEE and instead [Ca/Fe] ratios of Fornax stars are on average lower than those of both the LMg and HMg populations at all metallicities.  In the distribution of heavy elements, we find that the differences between Fornax and LMg stars in Ni abundances are similar to that in Mg ([Ni/Fe] is slightly lower in Fornax by about a tenth of a dex on average), whereas their Cr abundance distributions differ more significantly like Ca ([Cr/Fe] is lower in Fornax by a couple tenths of a dex on average).

In contrast to Fornax, giants from the more massive LMC (shown in Figure \ref{dsph}) typically have higher [X/Fe] ratios.  At metallicities \feh \ $\lesssim -1.0$, there is better agreement between the LMC giants and our LMg stars amongst their $\alpha$-element and Fe-peak abundances, e.g., the distributions of [Mg/Fe] and [Ca/Fe] ratios shown in the middle panels of Figure \ref{dsph}.  This may suggest that the metal-poor stars in our LMg population and the LMC have had analogous star formation histories, and ones that differ from both lower mass dSph satellites, and some of the more massive dSphs, such as Fornax.  At higher metallicities \feh \ $\gtrsim -1.0$, the LMC giants look like a chemical extension of the LMg stars.

Unfortunately, one of the most massive dSphs and therefore interesting satellites to compare with our Mg populations, the Sgr dSph, has been observed by APOGEE, but has been mostly excluded from our own sample by the stellar surface temperature restriction $T_{\rm eff} > 4000$ K.  So that we maintain as self-consistent a sample as possible, the latter requirement removes the coolest and brightest red giants from our sample, which have been analyzed by ASPCAP using a different grid of model atmospheres.  These infrared-bright stars, however are the only type of red giants that APOGEE has accessed and have data available to analyze in Sgr \citep[e.g.,][]{maj13,has17} because of the large distances to this dSph. For the same reason, but also because these younger stars are the dominant red giant population in the system, most other chemical abundance studies of the Sgr dSph also typically observe Sgr's more metal-rich stars \citep[e.g.,][]{sb07,has17}.  Nevertheless, while this younger, more metal-rich Sgr dSph population is not directly comparable to our more metal-poor populations, it has been noted to resemble a chemical extension of the low-$\alpha$ metal-poor stars in the MW \citep{has17} -- i.e., our LMg population.

A new study of the Sgr dSph, by \citet{muc17} appears to bear out this suggestion.  Using abundances of $\alpha$-elements measured from medium-resolution spectra, these authors show that both Sgr dSph and M54 stars (located at the center of the Sgr dSph) have similar $\alpha$-element chemical abundance patterns to LMC stars.  As may be seen in the bottom panels of Figure \ref{dsph}, and as is the case of LMC stars, there is an overlap in Mg and Ca abundances of the \citet{muc17} Sgr dSph and M54 stars with the LMg population (and to a smaller extent, the HMg population).

\subsection{Potential Origins}

As discussed above, the present dSph satellites of the MW typically have $\alpha$-element abundances that are too low to explain the origin of most MW stars observed by APOGEE \citep[and even the halo stars shown in][]{venn04,tol09}, at metallicities \feh \ $\gtrsim -1.5$, where we are interested in exploring the origin of the LMg and HMg populations.  This is demonstrated in our comparison with dSph stars in past studies and in our comparison with Fornax in Figure \ref{dsph}.  The one possible exception is the Sgr dSph, for which the dominant population looks like a possible metal-rich extension of the LMg population \citep{sb07,carretta10,has17}.  

Even if Sgr dSph stars may look more chemically similar to the LMg \citep[as is being revealed by larger samples that push to lower metallicities, see][]{muc17}, it seems unlikely that this particular satellite could have contributed the majority of the LMg stars.  This is evidenced by the full sky coverage of the LMg population with halo-like kinematics, whereas the Sgr dSph and its tidal tails are confined roughly to a plane in the sky \citep{maj03,lm10}.  Thus the majority of the LMg population (and HMg population, which has still higher $\alpha$-element abundances) does not seem to be accounted for by the accretion of dSph satellites like most of those around the MW now, particularly at higher metallicities, [Fe/H] $\gtrsim -1.5$.

As mentioned above, $\Lambda$CDM predictions, however, suggest that galaxies accreted earlier in the history of our Galaxy will tend to be more massive, resulting in higher \alphafe \ ratios than those being accreted today, for stars of the same metallicity \citep{font06,lee15}.  Additionally, cosmological hydrodynamical simulations including chemical evolution have reported complex scenarios where some of the accreted satellites could continue star formation activity in a bursty mode, producing stellar populations with a variety of levels of $\alpha$-element enrichment \citep[e.g.,][]{font06,tis12}.  The variation in the assembly histories of MW-mass galaxies has been shown theoretically to then shape the chemical patterns of their stellar halos \citep{font06,tis13}.

\citet{fa17}, using APOGEE data combined with distances, found that the innermost regions of the Galactic halo are dominated by stars with higher \alphafe \ ratios, but that dominance shifts to stars with lower \alphafe \ ratios at larger distances, at least for the moderately metal-poor regime probed by APOGEE (i.e., stars with [Fe/H] $\gtrsim -2$).  This [$\alpha$/Fe] variation supports the idea that more massive satellites with faster enrichment or star formation, and thus higher \alphafe \ ratios, may have been accreted earlier in the history of the MW to help form the inner regions of the halo.  

The lower \alphafe \ ratios of our LMg population compared to the metal-poor end of the thick disk, yet higher \alphafe \ ratios than current MW dSph satellites may then be evidence that these stars have been accreted from more massive dwarf systems early in the history of the MW.  Alternatively the LMg population may have originated from regions in the early MW halo with star formation similar to what would be expected in more massive dwarf galaxies.  It is interesting that the LMg population, a potentially accreted population, is a significant fraction of the metal-poor stars observed by APOGEE, at least between metallicities of about $-$1.5 and $-0.9$.

\citet{fis17} recently presented a study that examined neutron capture element abundances in stars selected from \citep{ns10}.  They found that in terms of light and heavy s-process elements ({\it ls} and {\it hs} respectively) the low-$\alpha$ halo stars have higher [{\it hs}/{\it ls}], which affirms results from \citet{ns11}, who found similar differences in [Ba/Y] (Ba is an {\it hs}- and Y an {\it ls}-element).  \citet{fis17} also found differences between these two populations in terms of their ratios of Y to Eu (an r-process element) and ratios of other s-process elements to Eu.  This is significant because the low [Y/Eu] ratios exhibited by the low-$\alpha$ halo stars, along with high [Ba/Y] ratios, are signatures seen in dSph stars, so that these neutron capture element patterns further support the accretion origin for the equivalent of our LMg population.

These conclusions are in agreement with those that have been drawn for the origin of low-$\alpha$ halo stars \citep{ns10,she12,haw15}, with which the LMg population seems to be associated.  In addition to exhibiting lower abundances of $\alpha$ and other elements \citep{ns10,ns11,she12,haw15}, low-$\alpha$ halo stars have been shown to have ages typically 2-3 Gyr younger than high-$\alpha$ halo and thick disk stars at any given metallicity, as well as larger orbital radii and distances from the Galactic mid-plane \citep{sch12}.  Additionally, \citet{sch12} found that the low-$\alpha$ halo stars they observed had larger eccentricities, clumped at values greater than 0.85 (i.e., $0.85 \lesssim e \lesssim 1.0$), whereas the observed high-$\alpha$ halo stars exhibit a greater spread in eccentricities ($0.4 \lesssim e \lesssim 1.0$).  The results of these and various other studies lend further support to the hypothesis that the low-$\alpha$ halo stars have been accreted.  

In contrast to the likely accretion origin for the LMg, the HMg populations's apparent net rotation and chemical similarity to the thick disk suggest an {\it in situ} formation similar or related to that of the thick disk.  If so, the HMg population might simply be a metal-poor extension of the thick disk, and the two may share an origin, whether through (1) dissipative collapse \citep{maj93}, (2) radial migration \citep[][note, however, that several recent simulations suggest that radial migration does not sufficiently heat the disk of the Galaxy --- cf. \citealt{min12,ver14}]{sel02}, or (3) by being ``kicked out'' or heated from initially colder orbits (possibly in the bulge or the colder disk) into more halo-like orbits by multibody encounters or the accretion of satellites \citep[possibly even those that contributed the accreted halo stars;][]{qui93,wal96,ns10,sch12,she12,jon16}.

Another possibility, proposed by \citet{haw15}, based on the chemical similarities between high-$\alpha$ halo stars and the thick disk, is that there may be a smooth transition between what they call the ``canonical halo'' and the thick disk, both of which they suggest formed {\it in situ}.  The HMg population  might then represent an intermediate, transitional stage between these two populations.  Because the HMg population has chemical abundance patterns similar to the thick disk, but with a lower apparent net rotation than the thick disk, it may be related to the MWTD reported by \citet{cb00} and \citet{bee02}.  If these two populations are the same, or are related, this would further support an {\it in situ} formation of the HMg population, as was proffered as the potential origin of the MWTD.  

The bifurcation of properties in metal-poor stars is reminiscent of the classic bimodality in Horizontal Branch (HB) types of the ``Younger Halo'' and ``Old Halo'' globular clusters between metalliticities $ -1.8 < $ [Fe/H] $ < -0.8$, which were thought to have been accreted from satellites and formed {\it in situ} respectively \citep{zinn93,zinn96}\footnote{At lower metallicities, [Fe/H] $ < -1.8$, \citet{zinn96} identifies a third group of metal-poor globular clusters that are spatially and kinematically distinct from the other two globular cluster populations, similar to the three-part division we found in metal-poor stars with the k-means clustering algorithm (Section 3.3).}.  While these globular cluster populations are no longer thought to be distinct in age alone \citep[due to complications in the differences between HB types;][]{gra10}, recent studies have found that there are two distinct age-metallicity relations amongst globular clusters \citep{van13,lea13,wag17} that cover similar age ranges, with the more metal-poor branch being about 2 Gyr younger than the metal-rich branch for a given metallicity.  In this paradigm, the more metal-poor and distant clusters are thought to have been accreted, whereas the more metal-rich clusters with more disk-like kinematics would have formed {\it in situ}.  With this picture of dual origins for globular clusters, it appears that both globular clusters and field stars separate consistently into {\it in situ} and accreted populations.

Putting the HMg and LMg populations within the context of prior studies of the thick disk and halo of the MW would benefit from the addition of full kinematics and spatial information for the APOGEE sample.  With the Gaia satellite about to deliver parallaxes and proper motions for stars at these magnitudes, this will soon be a reality.

\section{Conclusions}

We detect evidence for two distinct populations of metal-poor stars observed by APOGEE, discriminated by their \mgfe.  We study the chemistry and kinematics of these populations, and find multiple differences in their properties.  The separation between these populations in [Mg/Fe] is more pronounced for metallicities \feh \ $\gtrsim -1.5$.  While these populations are also distinguished by the patterns of other $\alpha$-elements, their distinctiveness is less apparent for heavier $\alpha$-elements such as Ca.  This variation in chemical separation may reflect some of the finer details of the differing nucleosynthetic processes forming these two populations such as the differential production and contribution of $\alpha$-elements in Type Ia supernovae or in Type II supernovae of different masses or metallicities \citep{tsu95,nom13}.  In addition to the $\alpha$-element differences, the LMg and HMg populations are distinct in their C+N, Al, and Ni abundances relative to Fe.  While our selection of the two populations used a by-eye discrimination in \mgfe-\feh \ space, we have also used two different clustering algorithms to search for distinct groupings in an 11-dimensional APOGEE chemical space.  Both of these methods generally reproduce our original selection and identify essentially the same two populations apparent in the [Mg/Fe]-[Fe/H] plane.  

We show that the LMg population exhibits halo-like kinematics, with little rotation and a large velocity dispersion of about 150-200 \kms.  The HMg population appears to be kinematically colder, with a rotational velocity $\sim$ 120-150 \kms \ and smaller velocity dispersion around 80-120 \kms.  This HMg population, however, includes some stars with radial velocities more consistent with halo-like orbits, similar to those found in other studies such as \citet{ns10}, and may reflect a chemical overlap between the LMg and HMg populations or a history tied to the formation of the thick disk (given the similarity between the chemistries of the HMg population and the thick disk).

Previous studies have also reported the detection of $\alpha$-element abundance differences in metal-poor stars, with some making selections specifically in Mg, as performed here \citep{ns10,nav11,ish12,she12,jj14,haw15}, albeit with fewer stars.  The advantage of our study is that we rely on a large sample of stars that have homogeneously determined abundances for many chemical species.  In addition, our sample is much larger in size, even at low metallicities, [Fe/H] $< -1.0$, where we have more than 1000 stars, which more than doubles the sample in \citet{jj14}, the largest of these studies.  Our analysis is of a sample that is free from kinematical biases, and probes a larger volume of the MW.  Finally, both by visual inspection and through the results of more sophisticated clustering algorithms, we are able to identify and separate the two distinct populations noted in past studies with greater statistical significance and reliability than before.

From the chemistry and kinematics of these two populations, we conclude that our LMg population is likely an accreted population of halo stars, formed in conditions similar to those in early dwarf galaxy satellites.  Examining the elemental abundance patterns of dSph stars \citep[from][]{venn04,let10,lem14}, we find that our LMg population stars have generally higher \alphafe \ ratios for stars with metallicities \feh \ $ \gtrsim -1.5$.  Thus it appears that if these stars (at least the more metal-rich LMg stars) were accreted earlier in the history of the MW, they were likely accreted from more massive satellites than present dSphs \citep{font06}.

Our HMg population, from its chemistry and its slow but significant net rotation, appears to contain mostly stars in the metal-poor end of the thick disk and/or may be related to the potentially distinct component of the MW, the MWTD \citep{cb00,bee02}.  Within this population, there are also stars that may have halo-like kinematics but chemistry similar to thick disk stars.  This would be consistent with the similarities between the \citet{ns10,ns11} thick disk and halo high-$\alpha$ stars, who suggest that the high-$\alpha$ halo stars (or equivalently our HMg stars with halo kinematics) may also be part of the dissipative component that also formed the thick disk.  This is similar to the picture presented by \citet{she12} who suggested that these stars could be {\it in situ} stars formed in such a dissipative collapse, or could be stars from the thick disk that were kicked into halo orbits.  The HMg population may then represent a combination of these possibilities.

Measuring more properties of the stars in these two populations may help us further distinguish them, provide more clues to their origins, and/or identify more sub-populations.  The origin of Eu in the low-$\alpha$ halo stars (LMg population) seen by \citet{fis17} is still not understood, and its relative abundance to s-process elements cannot be accounted for by the slow enrichment and low mass (1-3 M$_\sun$) AGB pollution \citet{fis17} authors use to explain the differences in {\it ls} and {\it hs} abundances in these stars.  Thus, as they suggest, further study of these populations with more r-process elements and larger samples may provide a better picture of the chemical evolution of metal-poor stars.


Expanded three-dimensional velocities would greatly expand our ability to study the kinematics and dynamics of the stars in these populations, but will require proper motions.  As suggested by \citet{nav11} and \citet{sch12}, the three-dimensional motions of stars in the low-$\alpha$ halo population provide evidence for accretion, so space motions would allow us to perform similar analysis with the populations seen in APOGEE.  Additionally, full space motions may help separate populations that have distinct kinematics but similar chemistry.  The physical distribution of the MW stars in our defined populations will be aided by incorporating accurate distance measurements \citep[an initial study of the distribution of metal-poor stars in APOGEE is given in][]{fa17}.  Finally, the companion paper by \citet[][Paper II in this series]{paper2}, further explores the chemical evolution of the two distinct metal-poor LMg and HMg populations identified in APOGEE.

\acknowledgements
The authors would like to thank the anonymous referee for her/his constructive comments and improvements, making this a better paper.  This research has made use of the Tool for OPerations on Catalogues And Tables \citep{topcat}.  CRH and SRM acknowledge National Science Foundation grants AST-1312863 and AST-1616636. CAP is thankful to the Spanish MINECO for funding for this research through the grant AYA2014-56359-P.  WJS wishes to thank the PAPIIT of México for financial support through project IN103014.  LC thanks the financial support provided by CONACyT of M\'{e}xico (grant 241732), by PAPIIT of M\'{e}xico (IG100115, IA101215, IA101517) and by MINECO of Spain (AYA2015-65205-P).  TCB acknowledges partial support from grant PHY 14-30152; Physics Frontier Center/JINA Center for the Evolution of the Elements (JINA-CEE), awarded by the US National Science Foundation.  J.G. F-T, DG, and BT gratefully acknowledge support from the Chilean BASAL Centro de Excelencia en Astrof\'{i}sica y Tecnolog\'{i}as Afines (CATA) grant PFB-06/2007.  DAGH was funded by the Ram\'{o}n y Cajal fellowship number RYC-2013-14182. DAGH and OZ acknowledge support provided by the Spanish Ministry of Economy and Competitiveness (MINECO) under grant AYA-2014-58082-P.  RRL acknowledges support by the Chilean Ministry of Economy, Development, and Tourism's Millennium Science Initiative through grant IC120009, awarded to The Millennium Institute of Astrophysics (MAS). RRL also acknowledges support from the STFC/Newton Fund ST/M007995/1 and the CONICYT/Newton Fund DPI20140114.

Funding for the Sloan Digital Sky Survey IV has been provided by
the Alfred P. Sloan Foundation, the U.S. Department of Energy Office of
Science, and the Participating Institutions. SDSS-IV acknowledges
support and resources from the Center for High-Performance Computing at
the University of Utah. The SDSS web site is www.sdss.org.

SDSS-IV is managed by the Astrophysical Research Consortium for the 
Participating Institutions of the SDSS Collaboration including the 
Brazilian Participation Group, the Carnegie Institution for Science, 
Carnegie Mellon University, the Chilean Participation Group, the French Participation Group, Harvard-Smithsonian Center for Astrophysics, 
Instituto de Astrof\'isica de Canarias, The Johns Hopkins University, 
Kavli Institute for the Physics and Mathematics of the Universe (IPMU) / 
University of Tokyo, Lawrence Berkeley National Laboratory, 
Leibniz Institut f\"ur Astrophysik Potsdam (AIP),  
Max-Planck-Institut f\"ur Astronomie (MPIA Heidelberg), 
Max-Planck-Institut f\"ur Astrophysik (MPA Garching), 
Max-Planck-Institut f\"ur Extraterrestrische Physik (MPE), 
National Astronomical Observatory of China, New Mexico State University, 
New York University, University of Notre Dame, 
Observat\'ario Nacional / MCTI, The Ohio State University, 
Pennsylvania State University, Shanghai Astronomical Observatory, 
United Kingdom Participation Group,
Universidad Nacional Aut\'onoma de M\'exico, University of Arizona, 
University of Colorado Boulder, University of Oxford, University of Portsmouth, 
University of Utah, University of Virginia, University of Washington, University of Wisconsin, 
Vanderbilt University, and Yale University.

\end{document}